\documentclass[%
 aip,
 amsmath,amssymb,
 reprint,%
,nofootinbib]{revtex4-1}

\usepackage{graphicx}
\usepackage{dcolumn}
\usepackage{bm}

\usepackage[utf8]{inputenc}
\usepackage[T1]{fontenc}
\usepackage{mathptmx}
\DeclareSymbolFont{epsilon}{OML}{cmm}{m}{it}
\DeclareMathSymbol{\epsilon}{\mathord}{epsilon}{"0F}
\usepackage{etoolbox}
\usepackage[dvipsnames]{xcolor}
\usepackage{graphicx} 
\usepackage{float}
\usepackage[english]{babel}
\usepackage{hyperref}
\usepackage{amssymb}

\def\sss{\scriptscriptstyle\rm}

\def\1s{_{1,\sss S}}
\def\2s{_{2,\sss S}}

\def\dulr{{\underline{\underline{\bf r}}}}
\def\dulR{{\underline{\underline{\bf R}}}}

\newcommand{\ben}{\begin{equation}}
\newcommand{\een}{\end{equation}}
\newcommand{\bea}{\begin{eqnarray}}
\newcommand{\eea}{\end{eqnarray}}

\makeatletter
\def\@email#1#2{%
 \endgroup
 \patchcmd{\titleblock@produce}
  {\frontmatter@RRAPformat}
  {\frontmatter@RRAPformat{\produce@RRAP{*#1\href{mailto:#2}{#2}}}\frontmatter@RRAPformat}
  {}{}
}%
\makeatother
\begin{document}

\preprint{AIP/123-QED}

\title{Different Flavors of Exact-Factorization-Based Mixed Quantum-Classical Methods for Multistate Dynamics}
\author{Evaristo Villaseco Arribas}
\affiliation{ 
Department of Physics, Rutgers University, Newark 07102, New Jersey, USA}
\email{evaristo.villaseco@rutgers.edu}
\author{Patricia Vindel-Zandbergen}
\affiliation{ 
Department of Chemistry, New York University, New York 10003, New York, USA}
\affiliation{ 
Department of Physics, Rutgers University, Newark 07102, New Jersey, USA}
\author{Saswata Roy}
\affiliation{ 
Department of Physics, Rutgers University, Newark 07102, New Jersey, USA}
\author{Neepa T. Maitra}%
 \affiliation{ 
Department of Physics, Rutgers University, Newark 07102, New Jersey, USA}
\email{neepa.maitra@rutgers.edu}

\date{\today}


\begin{abstract}
The exact factorization approach has led to the development of new mixed quantum-classical methods for simulating coupled electron-ion dynamics. We  
compare their performance for dynamics when more than two electronic states are occupied at a given time, and analyze: (1) the use of coupled versus auxiliary trajectories in evaluating the electron-nuclear correlation terms, (2) the approximation of using these terms within surface-hopping and Ehrenfest frameworks, and (3) the relevance of the exact conditions of zero population transfer away from nonadiabatic coupling regions and total energy conservation. Dynamics through the three-state conical intersection in the uracil radical cation as well as polaritonic models in one dimension are studied. 
\end{abstract}

\maketitle
\section{Introduction}
\label{sec:intro}
The accurate description of coupled electron-ion dynamics in photo-excited systems, is key to unravel the mechanisms underlying processes of chemical, physical and biological relevance. Some examples include photosynthesis~\cite{RNV17,TMF11,SGetal17}, radiation damage of DNA under UV light~\cite{Satzger10196,B815602F,RMGSG12} and charge dynamics in solar cell materials~\cite{AP14,CZNTP17,NSSKH17}. While a full quantum-mechanical treatment beyond Born-Oppenheimer (BO) is limited to a few degrees of freedom, mixed quantum-classical methods (MQC) provide an efficient way to simulate these processes while still recovering nonadiabatic effects~\cite{CB18}. The most commonly-used MQC schemes are Surface-Hopping (SH) and Ehrenfest (Eh)~\cite{T98,B11}, both of which involve propagating classical nuclear trajectories, but under two distinct forces, while coupled to the same equation for the quantum electronic evolution.
In Eh, the trajectories evolve under a mean-field force whereas in SH the the trajectories run on a single BO surface at each time, while stochastically and instantaneously hopping between them according to, usually, the fewest-switches scheme~\cite{T90}. Despite their simplicity, efficiency, and popularity, Eh and SH suffer from several well-known issues. While the mean-field nature of Eh precludes the possibility of describing wavepacket-splitting, SH is able to capture this. However, the disconnect between having coherent evolution of the electronic coefficients while the nuclear trajectories evolve on a BO surface at any given time, 
 leads to an internal inconsistency, responsible for over-coherence when the system evolves away from a region of electron-nuclear interaction. 
 Further, energy-conservation is imposed too strictly in both methods: at the individual trajectory level rather than conserving the energy over the ensemble representing the nuclear wavepacket~\cite{M16cc,M20}. In fact, in SH there is no unique way to conserve the energy~\cite{CGB17,B21,TSF21}, e.g. while velocity-adjustments along the direction of the nonadiabatic coupling (NAC) vector have been argued to be the most physical~\cite{Herman84,Pechukas69,CX95}, this method also leads to more forbidden hops, which exacerbates the internal consistency problem. Various fixes have been implemented in the electronic equation to account for decoherence effects, although their \textit{ad hoc} nature makes them not always accurate or reliable. Still, these methods have led to useful results in many situations, where the use of more sophisticated schemes is not feasible~\cite{CB18,TCVMB23,MMG20}. Recently these methods have been extended to the polaritonic regime, where the coupled electron-ion-photon dynamics of systems of thousands of molecules confined to an optical cavity can be described running nuclear trajectories on the hybrid photon-electronic (polaritonic) surfaces~\cite{LFTG17,FCPG20}.

An alternative view of nonadiabatic processes can be achieved through the exact factorization (EF) approach~\cite{AMG10,AMG12}. In EF, the time-dependent nuclear wavefunction evolves under a Hamiltonian of Schr\"odinger-form with a scalar and a vector potential that contain the exact electron-nuclear correlation, and depend on the time-dependent electronic wavefunction.  The electronic Hamiltonian in turn depends on the time-dependent nuclear wavefunction with terms embodying the exact correlation of nuclear motion on the electronic subsystem. 
The potentials in the nuclear equation lead to a uniquely determined force acting on the nuclei, which makes EF an ideal framework to develop MQC schemes. The resulting EF-based MQC methods contain extra terms in the electronic and nuclear equations compared to Eh or SH, which capture decoherence from first-principles~\cite{AMAG16,MAG15,MATG17,HLM18,AG21,PyUNIxMD}. A particularly significant effect of the EF term in these schemes was shown to arise in dynamics through the three-state conical intersection (CI) in a linear vibronic coupling model of the uracil cation~\cite{VMM22}: a SH scheme with the EF electronic equation accurately captured the reference MCTDH result, while traditional (decoherence-corrected) SH methods failed.
The analysis in Ref.~\cite{VMM22} suggested that generally when more than two BO states become occupied during the dynamics, that EF-based MQC methods would give improvements over the traditional MQC schemes because it includes quantum-momentum-driven transitions that are missing in the standard approaches. 

Here, we investigate this further, showing two distinct examples of multistate dynamics: one involving a three-state CI, and the other an avoided crossing where three pairs of states have appreciable pairwise couplings. We examine different EF-based MQC methods in each system, using coupled-trajectory schemes involving EF terms in both the electronic and nuclear equations, or in only the electronic equation with SH or Eh nuclei, as well as auxiliary-trajectory treatments of the EF terms that enable an independent-trajectory algorithm. We find that although the EF terms provide a new mechanism for population transfer particularly when more than two states are occupied, that  in some cases the effect is relatively small compared to the traditional terms, and the accuracy of the EF method depends also on the degree of violation of the exact conditions of zero net electronic population transfer away from NAC regions, and total energy conservation.

\section{Exact Factorization-based Mixed Quantum-Classical Approximations}
\label{sec:EF-based-mqc}
The coupled-trajectory mixed quantum-classical algorithm (CTMQC)~\cite{MAG15,AMAG16,MATG17,AG21} was derived by taking the classical limit of the nuclear EF equation, expanding the conditional electronic wavefunction in terms of BO states, and approximating some coupling terms as justified by their behavior in some model systems. The resulting CTMQC equations for the electronic coefficients  and nuclear force (where time-dependence is not explicitly indicated for simplicity of notation but assumed in all terms) associated with a given nuclear trajectory $\dulR^{(\alpha)}(t)$ in the ensemble are
\begin{equation}
\dot{C}_l^{(\alpha)}=\dot{C}_{l,Eh}^{(\alpha)}+\sum_\nu^{N_n}\sum_k\frac{{\bf Q}_\nu^{(\alpha)}}{M_\nu}\cdot\Delta{\bf f}_{\nu,lk}^{(\alpha)}\rho^{(\alpha)}_{kk}C_l^{(\alpha)}
\label{eq:el_eom}
\end{equation}
\begin{equation}
{\bf F}_\nu^{(\alpha)}={\bf F}_{\nu,Eh}^{(\alpha)}+\sum_\mu^{N_n}\sum_{l,k}\frac{2{\bf Q}_\mu^{(\alpha)}}{M_\mu}\cdot{\bf f}_{\mu,l}^{(\alpha)}\rho^{(\alpha)}_{ll}\rho^{(\alpha)}_{kk}\Delta{\bf f}_{\nu,lk}^{(\alpha)}
\label{eq:nuc_force}
\end{equation}
Both equations have an Ehrenfest-like term 
\begin{equation}
\dot{C}_{l,Eh}^{(\alpha)}=-iE_l^{(\alpha)}C_{l}^{(\alpha)}-\sum_{k}\sum_{\nu}^{N_n}{{\bf \dot{R}}}_{\nu}^{(\alpha)}\cdot{\bf d}_{\nu,lk}^{(\alpha)}C_{k}^{(\alpha)}
\end{equation}
\begin{equation}
{\bf F}_{\nu,Eh}^{(\alpha)}=-\sum_l\rho_{ll}^{(\alpha)}\nabla_{\nu}E_l^{(\alpha)}+\sum_{l,k}\rho_{lk}^{(\alpha)}\Delta E_{lk}^{(\alpha)}{\bf d}_{\nu,lk}^{(\alpha)}C_{k}^{(\alpha)}
\end{equation}
where ${\bf d}_{\nu,lk}^{(\alpha)}$ is the nonadiabatic coupling vector (NAC) along the $\nu$th nuclear coordinate 
between BO states $l$ and $k$ evaluated at the coordinate $\dulR^{(\alpha)}(t)$, and $\Delta E_{lk}^{(\alpha)}$ the BO energy difference between states $l$ and $k$.
The second terms in Eqs.(\ref{eq:el_eom}) and (\ref{eq:nuc_force}) are the corrections coming from EF with two key ingredients: the nuclear quantum momentum $\bf{Q}_\nu^{(\alpha)}$ evaluated at the position of the trajectory $\dulR^{(\alpha)}(t)$:
\begin{equation}
{\bf Q}_\nu^{(\alpha)}(t)=-\left.\frac{\nabla_\nu\vert\chi(\dulR)\vert^2}{2\vert\chi(\dulR)\vert^2}\right\vert_{\dulR = \dulR^{(\alpha)}(t)}\,,
\label{eq:qmom}
\end{equation}
and the time-integrated adiabatic force (an adiabatic momentum) accumulated on the $l$th surface, 
\begin{equation}
{\bf f}_{\nu,l}^{(\alpha)}=-\int_0^t\nabla_\nu E_l^{(\alpha)}d\tau\,,
\label{eq:accforce}
\end{equation}
with 
$\Delta{\bf f}_{\nu,lk}^{(\alpha)}={\bf f}_{\nu,l}^{(\alpha)}-{\bf f}_{\nu,k}^{(\alpha)}$. 

Although Eqs~(\ref{eq:qmom}) and (\ref{eq:accforce}) are definitions that emerge directly from the derivation of CTMQC, modified definitions have been proposed for both to ensure the algorithm satisfies some physical constraints. 
For the quantum momentum, the original definition ($\textbf{Q}_o$) that uses expression Eq.~(\ref{eq:qmom}) operates by reconstructing the nuclear density as a sum of Gaussians centered at the positions of the classical trajectories. A problem with the resulting algorithm, is that it can lead to {\it spurious transfer}~\cite{MATG17,VAM22}, meaning that when the ensemble of nuclear trajectories is in a region of negligible NAC, net population transfer over the ensemble can yet be induced, which is unphysical (that is, population transfers associated with different members of the ensemble should cancel).  The modified definition  of the quantum momentum, $\textbf{Q}_m$, is obtained by instead requiring the exact condition of zero net population transfer when the ensemble of trajectories experience zero NAC, fixing the spurious population transfer that might occur with $\textbf{Q}_o$. A deeper analysis on the effect of these two ways of computing the quantum momentum on the dynamics of model systems can be found in Ref.\cite{VAM22}. 

Regarding the integrated adiabatic force term, CTMQC with the original definition of Eq.~(\ref{eq:accforce}), and with either $\textbf{Q}_o$ or $\textbf{Q}_m$ for the quantum momentum, turns out to not always satisfy energy conservation. Instead, the modified algorithm, CTMQC-E, recently proposed in Ref.~\cite{VM23}, redefines this term to impose energy conservation over the ensemble.

Hence, a central feature of CTMQC is the coupling of the classical trajectories through the nuclear quantum momentum and integrated force terms. The coupling term induces electronic transitions that are needed to capture quantum (de)coherence effects. Further, the CTMQC electronic equation has been exploited in Eh and SH frameworks, either with coupled~\cite{GAM18,PA21,TLA22} or independent trajectories~\cite{HLM18,PyUNIxMD,KHM22}, leading to a family of EF-based MQC methods which are summarized  in Table~\ref{table:1}. We prefix the methods that use coupled trajectories to compute the quantum momentum by CT, while we use a suffix XF to indicate methods that compute this term via auxiliary trajectories in an independent trajectory scheme~\cite{HLM18}.  As previously discussed, the coupled trajectory methods compute the quantum momentum via coupled trajectories, either with $\textbf{Q}_o$ or $\textbf{Q}_m$, but the default is with $\textbf{Q}_m$, given its cure of the spurious transfer problem. 
In the auxiliary-trajectory methods, the quantum momentum is computed with the aid of virtual trajectories\footnote{Note that these trajectories are only used to approximate the quantum momentum and are not themselves trajectories of the ensemble that approximates the nuclear density} that are dressed by Gaussians of a chosen width parameter ($\sigma_\nu$) and locally construct the nuclear wavepacket associated with an independent trajectory enabling the computation of the quantum momentum using Eq.~\ref{eq:qmom}, which results in 
\begin{equation}
\mathbf{Q}_\nu^{(\alpha)}=-\frac{1}{2\sigma_\nu^2}\left(\mathbf{R}_\nu^{(\alpha)} -\sum_k\rho_{kk}^{(\alpha)}\mathbf{R}_{k,\nu}^{(\alpha)}\right)
\label{eq:auxqmom}
\end{equation}

These auxiliary trajectories are launched on non-active BO surfaces that become populated, and follow uniform velocity motion during each time interval. Their velocities during propagation are determined isotropically rescaling the velocity of the real trajectory enforcing energy conservation at each time-step. A key aspect is how to set their initial velocities. Some details related to some choices in propagating the auxiliary trajectories are discussed in Appendix~\ref{app:A}, and we will explore the effect of these different choices  on the dynamics of a model system in Sec.~\ref{sec:polariton}.

The independent (auxiliary) trajectories schemes are limited to approximating $\textbf{Q}_o$, since without knowledge of the ensemble it is not possible to compute $\textbf{Q}_m$. 
 While CTSH and CTEh utilize the electronic equation Eq.~(\ref{eq:el_eom}) in conjunction with nuclear dynamics determined by the usual fewest-switches SH scheme or the Eh force respectively, SHXF and EhXF are the analogous methods when using auxiliary trajectories to compute the quantum  momentum. CTEh has been explored for Tully model systems in Ref.~\cite{GAM18} (denoted there as CTMQCe) and CTSH was explored in Ref.~\cite{PA21}. The method we label here as MQCXF is when the EF terms are kept in both the electronic and nuclear equations with the use of auxiliary trajectories, i.e. MQCXF is used to denote the independent trajectory version of CTMQC. This was introduced in Ref.~\cite{HM22} (denoted there as EhXF); that work also explored additional implementations such as time-dependent Gaussian functions to construct the nuclear density or a modified expression of the accumulated force to ensure trajectory-wise energy conservation.
 
The independent trajectory approach, with auxiliary trajectories to mimic the coupling, reduces the computational cost enabling the simulation of large and complex systems; further computational expense is reduced also achieved by approximating the accumulated force (\ref{eq:accforce}) as the change in the momentum over a time-step at a given state. It was first introduced with SHXF in Ref.~\cite{HLM18} (originally called DISH-XF) and it has been applied to a range of light-induced processes on complex molecules~\cite{FMK19,FPMK18,FPMC19,FMC19}. 
In practise, the correction term derived from EF often gives similar results to decoherence-corrected SH schemes~\cite{VIHMCM21, MATG17,PyUNIxMD}, but one regime in which it gives a  qualitative improvement is when multistate CIs are involved. This was shown in Ref.~\cite{VMM22} for SHXF calculations of a model of the photo-excited uracil cation, and will be demonstrated in detail in the next section, where we will also compare with the other members of the EF-based MQC family.

\begin{table}[]
\centering
\begin{tabular}{cccc}
\hline
\multicolumn{4}{|c|}{\textbf{Coupled Trajectories EF-based MQC-methods}}                                                    \\ \hline
\multicolumn{1}{|c|}{Method} & \multicolumn{1}{c|}{Nuclear Force} & \multicolumn{2}{c|}{Quantum Momentum}                   \\ \hline
\multicolumn{1}{|c|}{\textbf{\textit{CTMQC}}}  & \multicolumn{1}{c|}{\textbf{\textcolor{OliveGreen}{Eh}}+\textbf{\textcolor{red}{XF}}}         & \multicolumn{1}{c|}{\textbf{\textcolor{Orange}{Q$_0$}}} & \multicolumn{1}{c|}{\textbf{\textcolor{Plum}{Q$_m$}}} \\ \hline
\multicolumn{1}{|c|}{\textbf{\textit{CTSH}}}   & \multicolumn{1}{c|}{\textbf{\textcolor{blue}{SH}}}            & \multicolumn{1}{c|}{\textbf{\textcolor{Orange}{Q$_0$}}} & \multicolumn{1}{c|}{\textbf{\textcolor{Plum}{Q$_m$}}} \\ \hline
\multicolumn{1}{|c|}{\textbf{\textit{CTEh}}}   & \multicolumn{1}{c|}{\textbf{\textcolor{OliveGreen}{Eh}}}            & \multicolumn{1}{c|}{\textbf{\textcolor{Orange}{Q$_0$}}} & \multicolumn{1}{c|}{\textbf{\textcolor{Plum}{Q$_m$}}} \\ \hline
\multicolumn{4}{l}{}                                                                                                        \\ \hline
\multicolumn{4}{|l|}{\textbf{Independent Trajectories EF-based MQC-methods}}                                                          \\ \hline
\multicolumn{1}{|c|}{Method} & \multicolumn{1}{c|}{Nuclear Force} & \multicolumn{2}{c|}{Quantum Momentum}                   \\ \hline
\multicolumn{1}{|c|}{\textbf{\textit{MQCXF}}}  & \multicolumn{1}{c|}{\textbf{\textcolor{OliveGreen}{Eh}}+\textbf{\textcolor{red}{XF}}}         & \multicolumn{2}{c|}{\textbf{\textcolor{Orange}{Q$_0$}}}                              \\ \hline
\multicolumn{1}{|c|}{\textbf{\textit{SHXF}}}   & \multicolumn{1}{c|}{\textbf{\textcolor{blue}{SH}}}            & \multicolumn{2}{c|}{\textbf{\textcolor{Orange}{Q$_0$}}}                              \\ \hline
\multicolumn{1}{|c|}{\textbf{\textit{EhXF}}}   & \multicolumn{1}{c|}{\textbf{\textcolor{OliveGreen}{Eh}}}            & \multicolumn{2}{c|}{\textbf{\textcolor{Orange}{Q$_0$}}}                              \\ \hline
\end{tabular}
\caption{EF-based MQC methods}\label{table:1}
\end{table} 
\section{Three-state conical intersections: Uracil cation}
\label{sec:uracil}
Our first example is the dynamics through the three-state CI in the photo-excited uracil radical cation, as modeled by an eight-mode linear vibronic coupling model that couples the four lowest states of the cation ($D_0$, $D_1$, $D_2$ and $D_3$); we refer the reader to Refs.~\cite{VMM22,AKM15} for the details of the model.\footnote{In the dynamics simulations, the parameters for the 2 modes of a$"$ symmetry enter only through the couplings, while we include 8 normal modes (6 a$'$ + 2 a$"$) to sample the initial conditions. Adding the harmonic and quartic terms from the a$"$ modes in our simulations do not alter the population dynamics} Ref.~\cite{VMM22} demonstrated that SHXF gave a significantly improved prediction for the population dynamics compared to Eh and to traditional decoherence-corrected SH methods, when beginning in the $D_2$ state.

\begin{figure}
\includegraphics[width=0.45\textwidth]{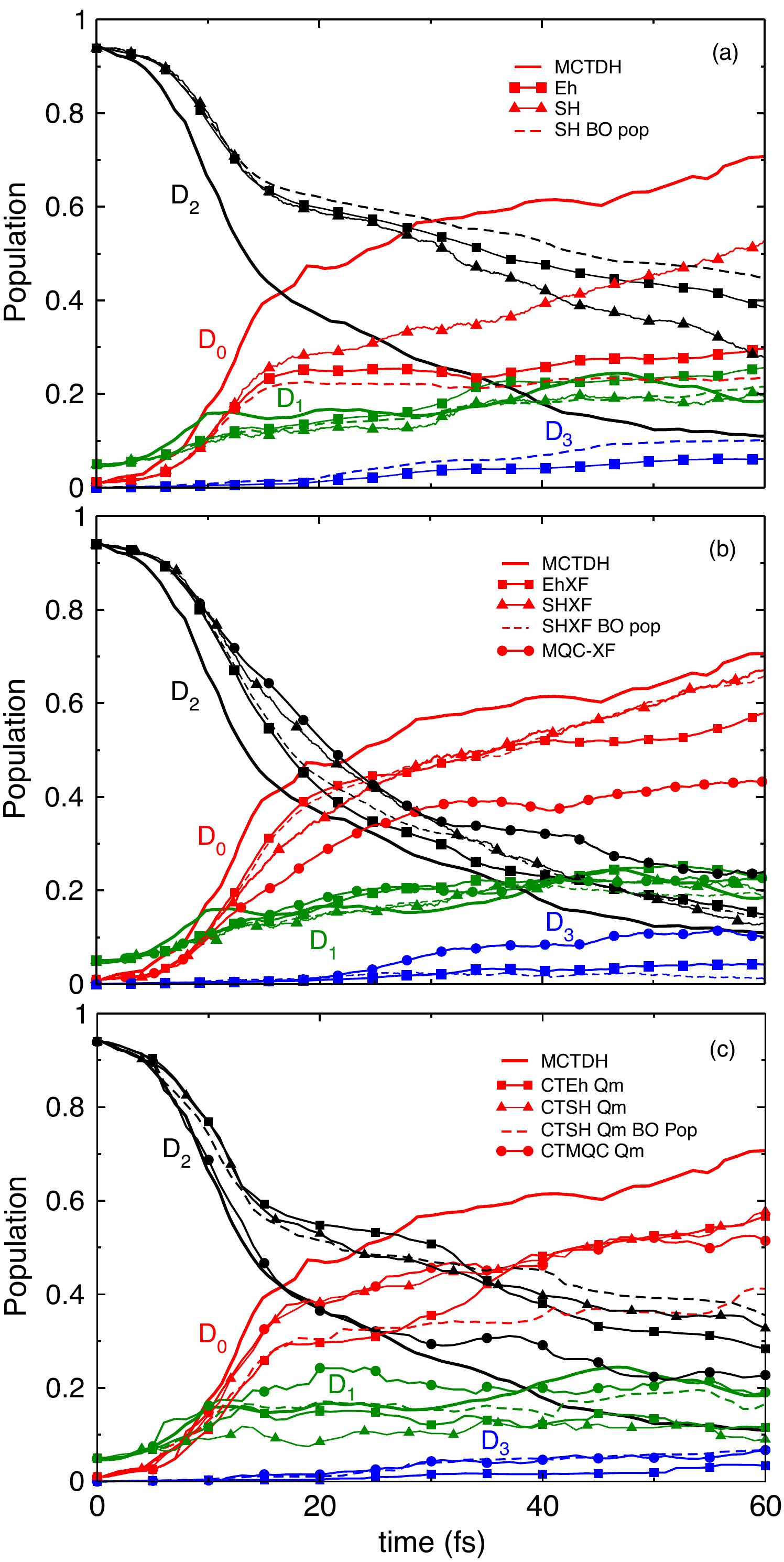}
\caption{Population dynamics in the uracil cation beginning in the mixed state with 94\% population in D$_2$, 5\% in D$_1$, and 1\% in D$_0$, along with the reference MCTDH (taken form ref.\cite{AKM15}) that begins in the diabatic state. Black, green, and red lines correspond to D$_2$, D$_1$, and D$_0$ states, respectively. (a) SH and Eh. (b) XF methods (c) CT methods with $Q_m$.}
\label{fig_D2_uracil}
\end{figure}

The EF term  in the electronic equation is crucial to describe the correct relaxation dynamics to the ground state through a three-state D$_0$/D$_1$/D$_2$ CI~\cite{AKM15,AWM16} as predicted by multiconfigurational time dependent hartree method (MCTDH)\cite{MMC90}: it induces electronic transitions mediated by the nuclear quantum momentum which significantly affect the electronic coefficients and hopping probabilities. 
Recalling Eq.~(\ref{eq:el_eom}), we observe that the 
 EF term has a distinct form from the usual \textit{ad hoc} decoherence corrections to SH: it depends on the nuclear quantum momentum and the accumulated force, and acts on all coefficients, coupling each of them to all occupied states. In contrast, the commonly used decoherence-corrections couple only an inactive state to the active state in a pairwise way~\cite{LAP16}. Besides that, the nonlinear dependence of the EF term on the coefficients gives a distinctly different dynamical behaviour than  the linear dependence of traditional decoherence correction schemes that underlies their character as simply a decoherence rate. This yields different
population dynamics than traditional methods where the difference is expected to be particularly significant in situations where more than two states become simultaneously associated with a given trajectory as it occurs during the passage through a three-state CI.
With the uracil cation studied in Ref.~\cite{VMM22}, SHXF was able to closely capture the population dynamics predicted by MCTDH where other traditional decoherence corrected SH methods fail (see also Fig.~\ref{fig_D2_uracil}). The number of net direct $D_2 \to D_0$ hops was found to be about twice as large in SHXF than the traditional methods after about $15$ fs, consistent with the faster relaxation to the ground state.
  
Here we consider the performance of the other members of the family of EF-based MQC methods of Table \ref{table:1}.
The independent trajectory calculations were performed using the \textit{}{PyUNIxMD} program package~\cite{PyUNIxMD}, whereas the coupled trajectory calculations using the \textit{G-CTMQC} package~\cite{GCTMQC}. Within the SH schemes, when a hop occurs, the momentum is rescaled along the direction of the NAC vector, and, if the hop is rejected (frustrated hop), we make the choice of keeping its direction rather than reversing it~\cite{JHT01,JST02,JT03}. 
Independent trajectory simulations require a time step of $dt=0.1$ fs and $N_{tr}=$1000 trajectories for convergence. Coupled trajectory calculations require a smaller time step of $dt=0.001$ fs but reach convergence with fewer trajectories; here results with $N_{tr}=$400 trajectories are shown. 
We employ the same initial conditions as in Ref.~\cite{VMM22}, sampled from a Wigner distribution of the neutral uracil ground state (S$_0$) equilibrium geometry with variances obtained from the frequencies of the eight modes in the model. 
 To compute the quantum momentum, in the independent-trajectory methods we used a fixed width of $\sigma= 0.08$ a.u. for the Gaussians centered on the auxiliary trajectories on each degree of freedom; this number is determined from the average of the initial distribution of the nuclear trajectories of the C=C, C=O, and C-N bonds. 

Figure \ref{fig_D2_uracil} shows the population dynamics starting on the D$_2$ state,  computed from the traditional methods Eh and SH in the top panel, the EF-based independent-trajectory methods SHXF, EhXF, and MQCXF in the middle panel, and the EF-based coupled-trajectory methods CTMQC, CTEh and CTSH  in the bottom panel, all including the MCTDH as the benchmark reference. Since the initial MCTDH state is a diabatic state with  94\% population on the D$_2$, 5\% on D$_1$ and 1\% in D$_0$ adiabatic states, we approximate the initial state in the independent-trajectory calculations via a statistical mixture by distributing the trajectories among the states accordingly. On the other hand, for coupled trajectory schemes each trajectory is initialized in a pure state, that is in a superposition of eigenstates, with the modulus-square of the coefficients matching the MCTDH initial adiabatic populations. This is a more faithful representation of the initial electronic state of the MCTDH simulation at each nuclear configuration than the incoherent statistical mixture, and it has an important immediate effect on the dynamics in coupled-trajectory schemes: 
the quantum momentum couples the trajectories from the start, in contrast to the initial mixed state where only one coefficient being non-zero sets the XF terms in Eqs.~(\ref{eq:el_eom}-\ref{eq:nuc_force}). In our simulations however, we do not have access to the initial exact complex coefficients from MCTDH, and  only have the populations. So,  we have simply chosen them to be real in Fig.~\ref{fig_D2_uracil}, and we will return to the impact of this choice, and the comparison with using an incoherent statistical state, shortly.

For the SH schemes, both the trajectory-average of the electronic populations $|C_l(t)|^2$ and the fraction of trajectories running on each state, $\Pi_l(t)=N_l(t)/N_{tr}$ are shown. We observe in the top panel, that after  the initial decay of the D$_2$ and simultaneous rise of D$_1$ and D$_0$ populations,  the $D_2$ population predicted by SH and Eh shoulder off, deviating from the continued decay predicted by MCTDH. SH suffers from overcoherence, as the number of trajectories running on each state differs from the electronic populations, and although the traditional energy-based decoherence-corrected scheme~\cite{GP07,GPZ10} corrects this, it does not improve the qualitatively wrong dynamics after 15 fs, as shown in Ref.~\cite{VMM22}. (Interestingly, the augmented FSSH scheme~\cite{SOL13,JAS16} neither cures the internal consistency~\cite{VMM22}, nor improves the population behavior). While the electronic populations are similar for Eh and SH,  SH with the fraction of trajectory measure performs better than Eh, showing a faster decay of D$_2$ and $\sim$20\% more population transfer to D$_0$. We observe that while the Eh and SH electronic populations yield some transfer to the third excited cationic state $D_3$, unlike in the MCTDH reference, there are no hops to this state, so again the fraction of trajectory measure in the SH calculation captures this aspect better. 

Turning now to the middle panel, we observe that the EF correction used in the independent trajectory methods gives a significant improvement to the dynamics. 
 When applied to just the electronic equation in the SH scheme,  SHXF, as observed in Ref.~\cite{VMM22}, more faithfully reproduces the MCTDH populations than the traditional methods. Note that whereas in SH the electronic evolution is usually viewed as merely a support for the trajectory propagation, in SHXF the electronic equation plays a key role since the equation is derived from the rigorous EF method. The electronic populations of SHXF are closer to that of EhXF at intermediate times than to the fraction of trajectories measure in SHXF but internal consistency is recovered at longer times; the violation at intermediate times is coming from trajectories that are initially on the states  D$_0$ and D$_1$, and can be compared with the very close internal consistency  observed in Ref.~\cite{VMM22} for dynamics beginning in the 100\% adiabatic state. 
The electronic populations measure show  less population transfer to $D_3$ than the traditional methods, and even this correctly becomes negligible when the fraction of trajectories measure is used . We see here, that even if the EF correction is applied only to the electronic equation in the Eh scheme in EhXF, the populations are also well-reproduced; there is a larger underestimation of $D_0$ at longer times at the expense of some population in $D_3$. Applying the EF correction  also to the nuclear force in MQCXF, the results are a little worse than in EhXF and SHXF, although still the trends are much better than in Eh and SH, and we conjecture that this may be due to a violation of energy-conservation~\cite{HM22}, as evidenced by the larger population in $D_3$ and smaller population in $D_0$. The larger population in the $D_3$ state of MQCXF with respect to the also non energy-conserving EhXF~\cite{VM23}, suggests the main contribution to the energy violation in this system comes from $\mathbf{F}_{XF}^{(\alpha)}$ rather than from $\dot{\rho}_{XF}^{(\alpha)}$.

The lowest panel in Fig.\ref{fig_D2_uracil} shows the results when the quantum momentum is computed from coupled trajectories. CTMQC improves the population behavior compared to its independent-trajectory analogue MQCXF, and very closely follows the initial MCTDH D$_2$ population decay up to around 30 fs, deviating from the MCTDH reference after 30 fs, possibly due to energy non-conservation~\cite{VM23}. Interestingly the increase in the population of the $D_3$ state in CTMQC is smaller than in MQCXF, potentially suggesting that energy conservation might be ameliorated using coupled trajectories, although further investigation is needed. The underestimate of the increase of the D$_0$ population relative to the MCTDH reference comes mainly from an overestimation of the D$_2$ to D$_1$ population transfer. On the other hand for CTEh and CTSH, unlike for their independent trajectory counterparts, there is a smaller improvement over traditional Eh or SH especially for the D$_2$ population behavior, and once again we see this population incorrectly shouldering off at around 15 fs. It appears that the EF contribution to the nuclear force is key to capture the faster and continuing D$_2$ population decay  for the CT methods, while this additional force appears less important in the independent-trajectory XF methods. 

\begin{figure}
\includegraphics[width=0.45\textwidth]{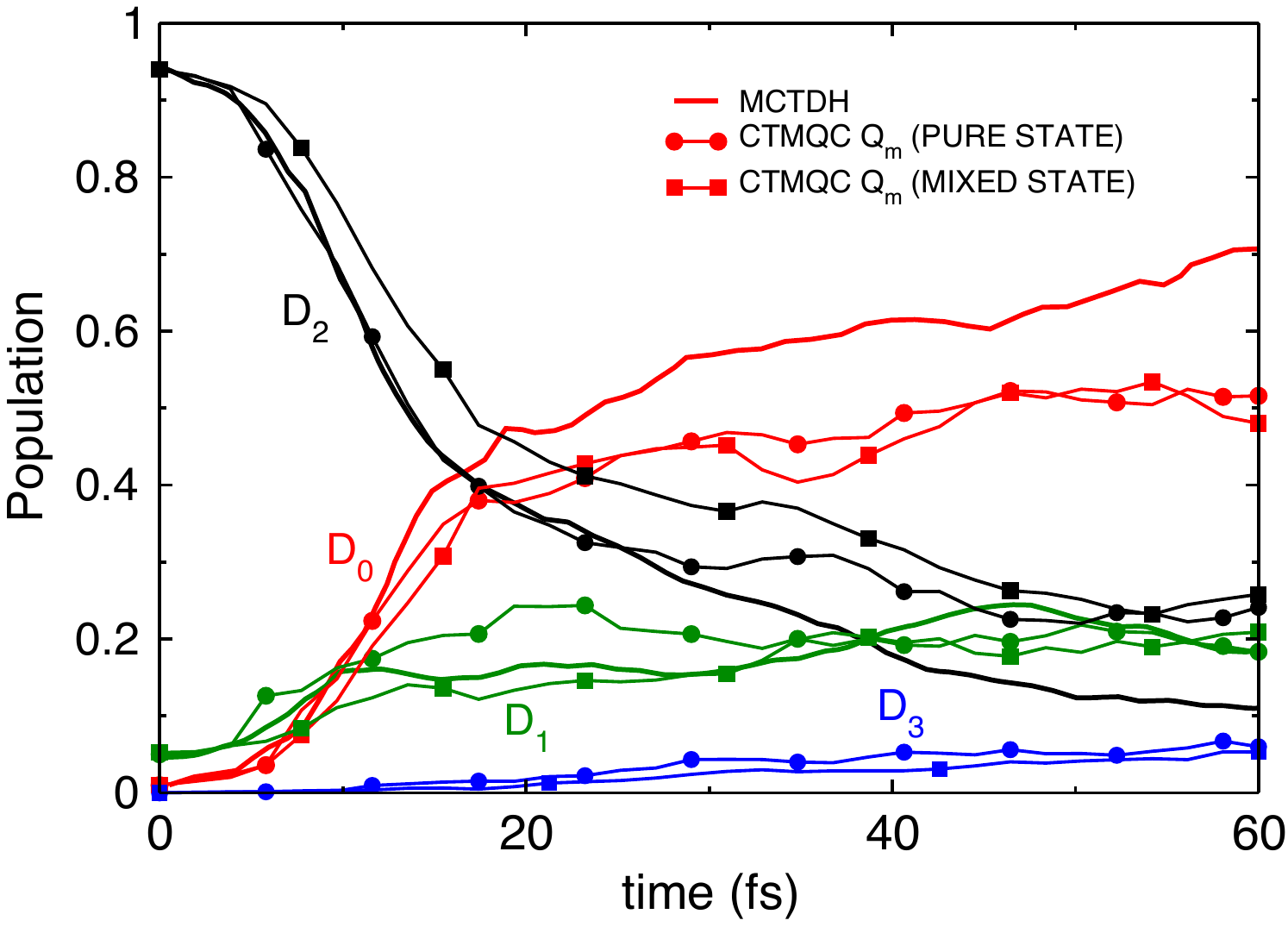}
\caption{Population dynamics computed from CTMQC with $Q_m$ in the uracil cation beginning in a mixed incoherent state (squares), and a pure coherent state with real electronic coefficients (circles) together with the MCTDH reference (solid lines).}
\label{fig:dif_init}
\end{figure}

As mentioned earlier, the initial electronic state in the CT calculations was chosen to be a pure state for each trajectory, with real coefficients whose square equals the adiabatic populations. To get a sense of the impact of this, Figure~\ref{fig:dif_init}
compares the differences in the CTMQC dynamics initializing the ensemble in the incoherent mixed state with electronic density operator $\hat{\rho}=0.01\vert D_0 \rangle\langle D_0 \vert+0.05\vert D_1 \rangle\langle D_1 \vert+0.94\vert D_2 \rangle\langle D_2 \vert$ where 1$\%$ of the trajectories are initialized with $\rho_{D_0}=1$, 5$\%$ with $\rho_{D_1}=1$ and 94$\%$ with $\rho_{D_2}=1$ (as was done in the independent-trajectory calculations)  versus a pure state where all trajectories in the ensemble are in the same coherent linear superposition state $\vert\Phi_{(0)}\rangle=\sqrt{0.01}\vert D_0\rangle+\sqrt{0.05}\vert D_1\rangle+\sqrt{0.94}\vert D_2\rangle$ (chosen with real phases). Figure \ref{fig:dif_init} shows the differences in the CTMQC electronic populations in the incoherent mixed state versus this pure state with real coefficients. As we can observe an active quantum momentum is needed to capture the fast initial decay in the $D_2$ population. Further investigation on the impact of the different initializations and effect of the phases on the coupling terms of EF-based trajectory schemes is currently underway.

\begin{figure}
\includegraphics[width=0.45\textwidth]{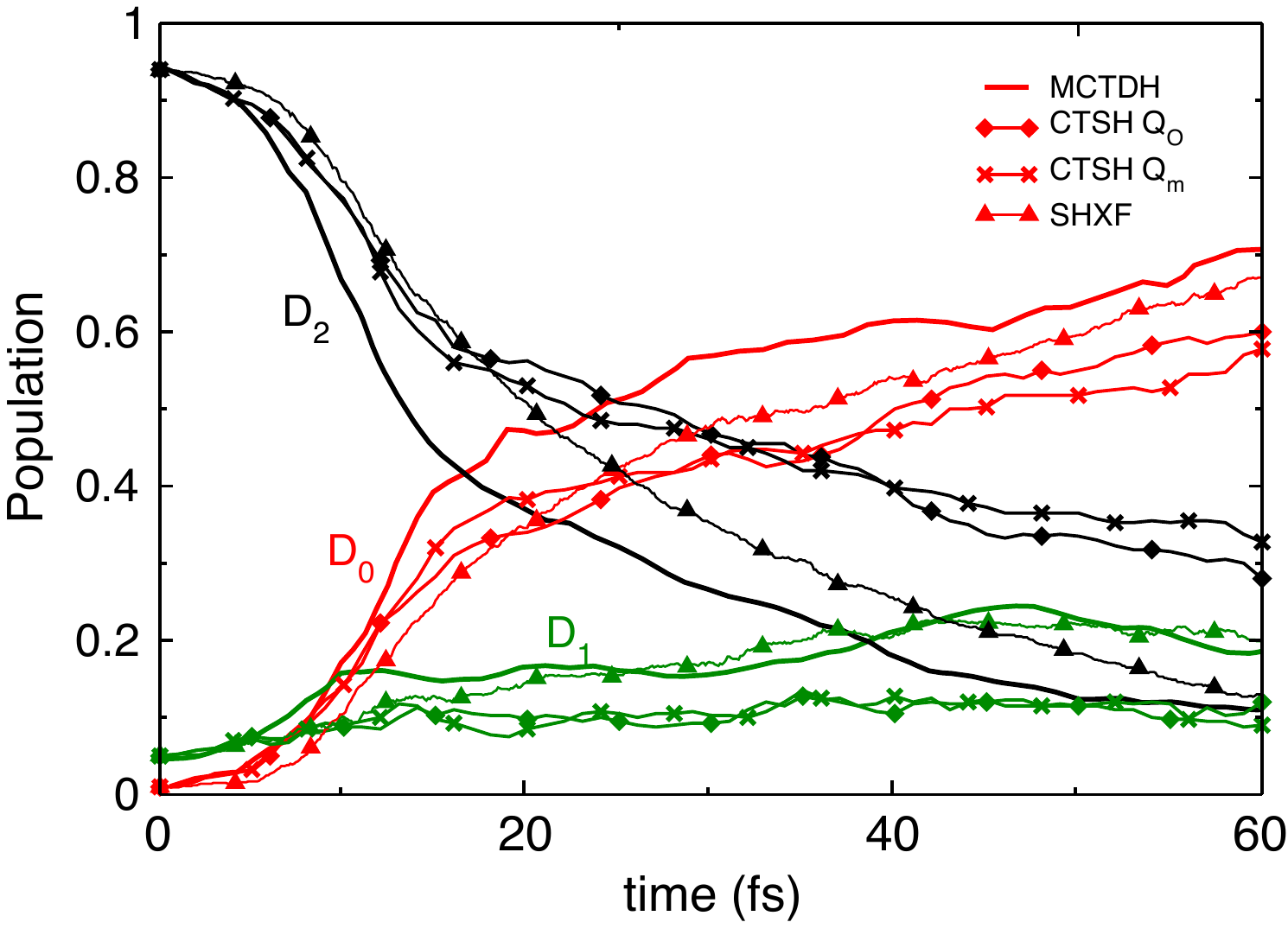}
\caption{CTMQC population dynamics and CTSH population dynamics (dashed lines) and fraction of trajectories in the uracil cation beginning in the mixed state with 94\% population in D$_2$, 5\% in D$_1$, and 1\% in D$_0$, along with the reference MCTDH reference. Black, green, and red lines correspond to D$_2$, D$_1$, and D$_0$ states, respectively. Results from $Q_o$ and $Q_m$ definitions are compared.}
\label{fig_D2_sh}
\end{figure}


Perhaps most puzzling is why CTEh and CTSH perform worse than EhXF and SHXF, and why the EF correction appears to have a much smaller impact in these two algorithms for this system when computed in the coupled trajectories scheme than in the independent trajectory scheme. The coupled-trajectory methods utilize the $\textbf{Q}_m$ definition to automatically satisfy the condition of zero net transfer in regions away from a NAC, while the independent-trajectory methods are based on $\textbf{Q}_o$, as discussed earlier. A natural question then arises:  if we perform the coupled-trajectory dynamics computing the quantum momentum with $\textbf{Q}_o$, would they be closer to their independent-trajectory EF counterpart   Fig~\ref{fig_D2_sh} compares the results of a CTSH calculation using $\textbf{Q}_o$, and shows that it predicts the D$_0$ population slightly closer to SHXF than when using $\textbf{Q}_m$, although the D$_2$ population decay  still shoulders at around 15 fs, and transfer to D$_1$ is underestimated. Even though the same equations underlie both SHXF and CTSH with $\textbf{Q}_o$, the locality of the auxiliary trajectories when launched for each independent trajectory can make the action of the quantum momentum quite different from when it is computed with coupled trajectories. In SHXF, auxiliary trajectories  are launched on inactive surfaces to compute the quantum momentum when the electronic population on the inactive state becomes larger than a small threshold. 
The auxiliary trajectories start at the same position as the actual independent trajectory and begin to deviate from the real trajectory in nuclear space due to the different adiabatic forces, thus giving a net contribution to the quantum momentum before decoherence fully sets in. The situation is quite different for the quantum momentum $\textbf{Q}_o$ computed from coupled trajectories, where each trajectory is, in principle, coupled to the entirety of the trajectories in the ensemble, contributing in a more non-local way to the quantum momentum as soon as an inactive state acquires a non-zero population. This is consistent with the faster decay at short times ($<5$ fs) in CTSH with $\textbf{Q}_o$ (and with $\textbf{Q}_m$) than in SHXF. In CTSH, even trajectories on the same surface as the active state contribute whether they may be close to the current position of the trajectory or far, having had different histories, unlike with auxiliary trajectories which are launched only on the inactive surfaces.
 At longer times, it can happen that contributions to the quantum momentum from trajectories exploring diverging paths in the nuclear space may have canceling effects, leading to only a very small correction over SH; and a more detailed investigation of this is left to future work.

\begin{figure}
\includegraphics[width=0.45\textwidth]{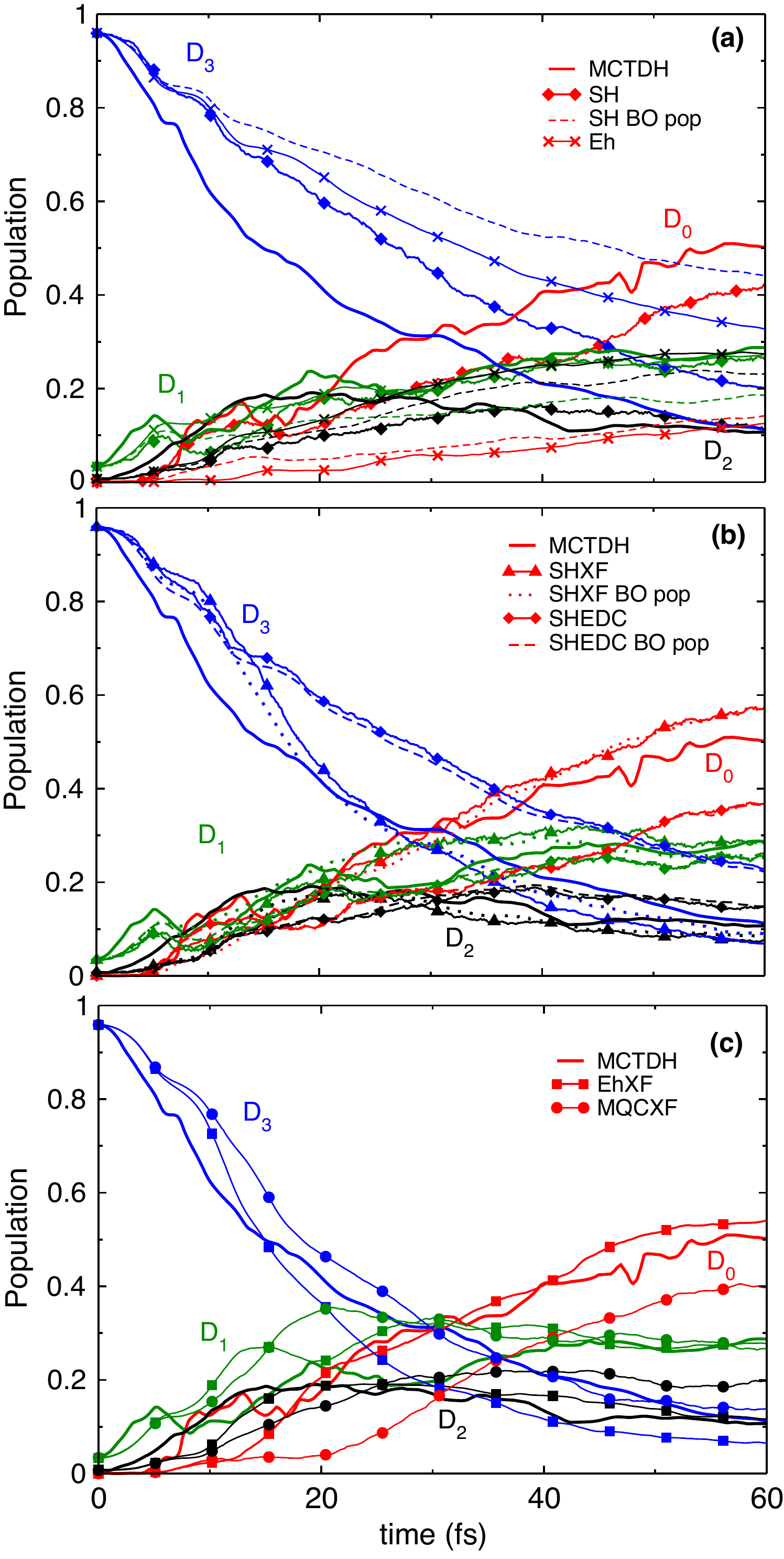}
\caption{Population dynamics in the uracil cation beginning in the mixed state with 97\% population in D$_3$, 0.7\% in D$_2$, and 3.3\% in D$_1$, along with the reference MCTDH (taken form ref.\cite{AKM15}) that begins in the diabatic state. Black, green, and red lines correspond to D$_2$, D$_1$, and D$_0$ states, respectively. (a) Ehrenfest population and SH electronic population (dashed lines) and SH fraction of trajectories. (b) Populations as fraction of trajectories calculated from  SHXF and SHEDC. (c) Population dynamics computed with EhXF and MQCXF.}
\label{fig_D3_uracil}
\end{figure}

We next consider dynamics starting on the D$_3$ state, shown in Figure \ref{fig_D3_uracil}, where again the D$_0$/D$_1$/D$_2$ three-state intersection plays a crucial role and focus here on the independent trajectory  EF-based MQC methods.  
The same 1000 initial conditions as for the dynamics starting on D$_2$ are used for the independent-trajectory calculations, again with a time step of $dt=0.1$ fs and a fixed $\sigma=0.08$ a.u. 
The initial state is prepared with populations identical to the initial diabatic state used in the MCTDH reference~\cite{AKM15}, which corresponds to a distribution of 96\% of trajectories in D$_3$, 0.7\% in D$_2$ and 3.3\% in D$_1$. 
As evident from Fig.~\ref{fig_D3_uracil}, in the reference MCTDH calculation, the D$_3$ state decays with population transferred initially to the D$_1$ and D$_2$ states, with D$_0$ beginning to be populated after a short time ($\sim5$ fs). Up to 15 fs, the population in three states, D$_2$, D$_1$ and D$_0$ rises simultaneously, before D$_2$ levels off while D$_1$ and D$_0$ steadily increases, with D$_0$ showing a larger slope. At 60 fs the population is distributed amongst the 4 states, with 50\% population in D$_0$, 30\% in D$_1$ and 20\% equally distributed between D$_3$ and D$_2$. 
The traditional SH and Eh methods shown in the top panel again show the D$_3$ population decays much slower and the population transfer is underestimated. The fraction of trajectory measure of population in SH is closer to MCTDH than Eh populations, although the internal inconsistency is large.
The middle panel shows the trajectory-average electronic populations and the fraction of trajectories for the energy-based decoherence correction method~\cite{GPZ10,GP07}, SHEDC, and SHXF,   compared to the reference MCTDH. 
SHEDC and SHXF correct the internal inconsistency of the SH method, with SHXF predicting dynamics closer to MCTDH overall. The EDC correction has little effect other than correcting the internal inconsistency, and it may even worsen population trends. On the other hand, although initially slower, the SHXF decay-rate of the D$_3$ state is too fast between about  15 fs - 25 fs, and ultimately overestimates the population transfers to D$_1$ and D$_0$, but overall is more accurate over the range shown than SHEDC.  None of the corrections can capture correctly the initial increase of D$_2$, with SHXF performing slightly better at later times. Turning now to the lower panel (Fig.~\ref{fig_D3_uracil}(c)),  EhXF and MQCXF correct the dynamics predicted by Eh, improving the initial D$_3$ decay. MQCXF follows MCTDH quite closely, while, like SHXF, EhXF overestimates this population transfer from $\sim$15 fs. Again, neither of the methods capture correctly the initial D$_2$ nor the D$_0$ rise, but EhXF catches up to agree with MCTDH from $\sim$15 fs. The initial transfer to
D$_1$ is overestimated in both methods, however, at later times, matches the MCTDH prediction well. 

The generally good performance of the independent trajectory EF-based methods SHXF, EhXF (and MQCXF to a lesser extent) for this system suggest that the dynamics is in a regime where effects of the spurious transfer are relatively minor compared to the overall correction of the population trend compared to the traditional methods. Energy non-conservation may be a reason for the slight underperformance of MQCXF and CTMQC~\cite{VM23}, and further study into this is needed. The original definition of the quantum momentum when evaluated with auxiliary trajectories appears to be effective in capturing quantum-momentum driven electronic transitions that are missing in standard methods, and the impact of these transitions can be more important than their violation of the spurious transfer condition. When using coupled trajectories, the non-locality of these terms yields significant differences. Specifically with coupled-trajectories, both quantum momentum-driven electronic transitions and the quantum momentum term in the force, as in CTMQC, are needed to accurately capture the population dynamics for this system.
 
\section{Three-state interactions: Model of Polaritonic Chemistry}
\label{sec:polariton}
As demonstrated in Sec.~\ref{sec:uracil} and Ref.~\cite{VMM22} for the dynamics of the uracil radical cation through a 3-state CI, the EF terms provide a significant correction to the standard methods in situations where more than two electronic states are occupied at a given time and nuclear configuration. Another case in which we expect a similar behavior, with several electronic states associated with a given trajectory, are polaritonic systems. In polaritonic systems, a molecule confined to an optical cavity couples strongly with vacuum fluctuations of the confined radiation field forming hybrid light-matter states known as polaritons. This light-matter coupling distorts the (BO) energy landscape of the molecule resulting in altered photochemical dynamics
~\cite{RMDC18,FGF22}. The density of states is increased with respect to the cavity-free situation as a consequence of the molecular interaction with the confined light modes, thus we expect an increased number of avoided crossings and multi-state intersections that the nuclear trajectories will encounter in their evolution.

\begin{figure}
	\includegraphics[width=0.45\textwidth]{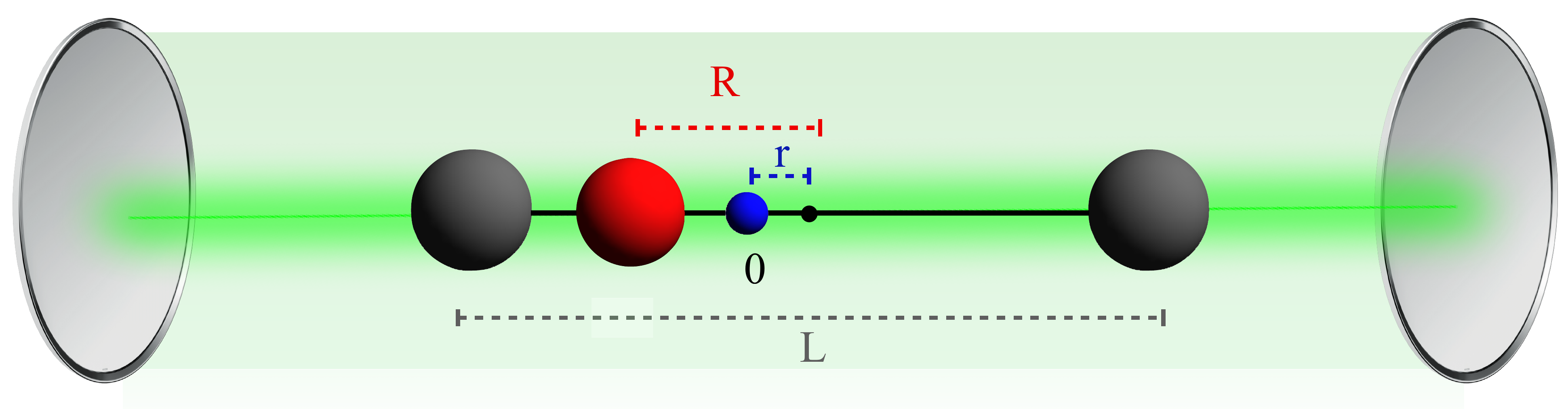}
	\caption{Shin-Metiu model in an optical cavity.}
 \label{fig:SMfigure} 
\end{figure}

In this section we study a one-dimensional system, a Shin-Metiu model~\cite{SM95,FHS97,YHS97} placed in an optical cavity coupled to one photon mode. This model, consists of one electron and one ion which are allowed to move, along with two fixed ions separated a distance $L$ (see Fig~\ref{fig:SMfigure}). In the non-relativistic limit and the long-wavelength approximation, the total Hamiltonian for this system has the form~\cite{FARR17,RTFHR18,HARM18,T13,RFPATR14}
\begin{equation}
	\hat{H}(R,r,q) = \hat{H}_m(R,r) + \hat{H}_p(q) + \hat{V}_{pm}(R,r,q) \\
	+ \hat{V}_\textrm{SP}(R,r,q)
 \label{eq:Hpol}
\end{equation}
where $r$, $R$, and $q$ indicate the electronic, nuclear and photonic degrees of freedom respectively. The first term is the matter Hamiltonian $\hat{H}_m = \hat{T}_n + \hat{H}_{\rm BO}$, where $\hat{T}_N=-\frac{1}{2M}\frac{\partial^2}{\partial R^2}$ is the nuclear kinetic energy operator, and $\hat{H}_{\rm BO}$ the BO Hamiltonian, which for our Shin-Metiu model reads
\begin{equation}
\hat{H}_{BO}=-\frac{1}{2}\frac{\partial^2}{\partial r^2}+ \sum_{\sigma=\pm1} \Big(\frac{1}{|R+\frac{\sigma L}{2}|} - \frac{\textrm{erf}(\frac{|r + \frac{\sigma L }{2}|}{a_\sigma})}{|r+\frac{\sigma L}{2}|}\Big) 
		+ \frac{\textrm{erf}(\frac{|R -r|}{a_f})}{|R-r|},
  \label{eq:SM}
\end{equation}
here $\textrm{erf}$ is the Gauss error function and $M$ the mass of the proton ($1836$ a.u.). The model parameters $a_{+1}$, $a_{-1}$, $a_f$ and $L$ will be tuned appropriately to manufacture a 3-state crossing region. The photonic Hamiltonian reads $\hat{H}_p = \frac{1}{2}(\hat{p}^2 + \omega^2 \hat{q}^2)$
where the frequency of the photon mode is  $\omega$, and $\hat{q} = \sqrt{1/2\omega}(\hat{a} + \hat{a}^\dagger)$
 is the electric displacement operator with the conjugate variable  $\hat{p}$  proportional
to the magnetic field. The light-matter coupling has bilinear form $\hat{V}_{pm} = \omega\lambda\hat{q}(R-r)$, and the self-polarization term
term  $\hat{V}_\textrm{SP}=\frac{1}{2}[\lambda(R-r)]^2$ 
which is often negligible for a single cavity mode.
Although the matter–photon coupling strength $\lambda$ is generally proportional to the mode function
of the cavity, here we take it constant assuming that the cavity
length is much larger than the distance between the two fixed ions $L$. The polaritonic surfaces, i.e. the eigenvalues of $\hat{H}-\hat{T}_n$, play the role that the BO surfaces play in the cavity-free case~\cite{GGF16}, providing the playground for the nuclear motion. 

This model was used in Refs.~\cite{LHM19,HLRM20,MRHLM21} to study a proton-coupled electron-transfer (PCET) process. Partial suppression of the PCET reaction was observed due to the cavity-coupling; part of the wavepacket becomes trapped in a local potential well in one of the polaritonic surfaces that is induced by the cavity-coupling, and is unable to reach the region of electron-nuclear interaction. This suppression reaction was found to be strongly affected by the number of photon modes included in the cavity. Ref.~\cite{{MRHLM21}} studied the cavity-induced suppression of the PCET and its dependence on the initial state. It was found that, unlike the polaritonic surfaces, the structure of the exact time-dependent potential energy surface from the EF approach correlated directly with the proton dynamics. Ref.~\cite{HMWH22} also studied this polaritonic system with nonadibatic mapping approaches, showing they outperform traditional SH and Eh. 

\begin{figure}
	\includegraphics[width=0.45\textwidth]{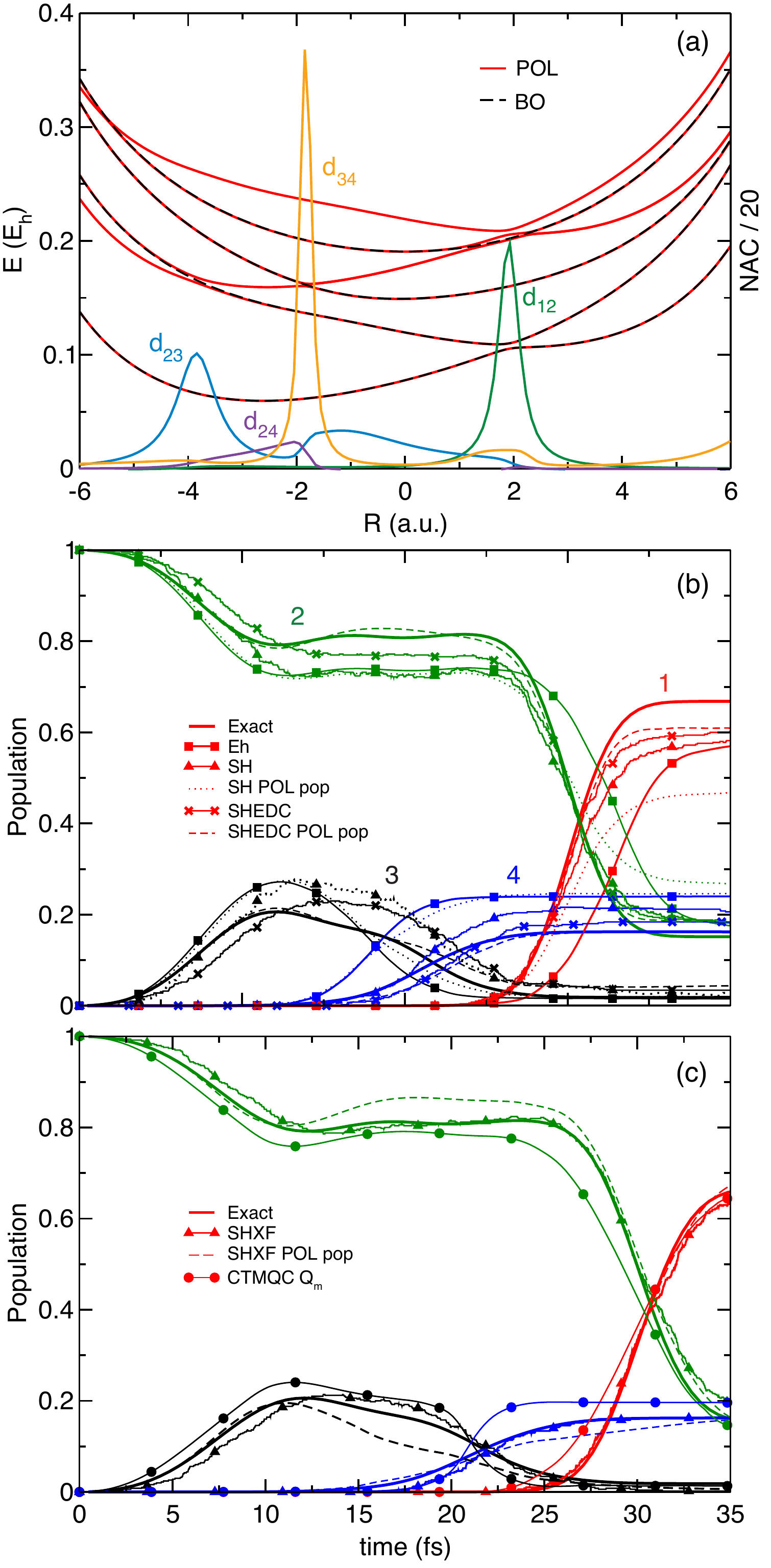}
	\caption{a) Polaritonic surfaces, BO PESs and NACs for Shin-Metiu model in cavity with
	$\lambda = 0.005$, as studied in Ref.~\cite{LHM19,HLRM20,MRHLM21,HMWH22}. (b) Eh populations on the polaritonic surfaces and SH and SHEDC populations and fraction of trajectories. (c) SHXF populations and fraction of trajectories, and CTMQC with $\textbf{Q}_{m}$ electronic populations.}
 \label{fig:LHM19merge} 
\end{figure}

We first consider the same set of parameters that represent the PCET model of Refs.~\cite{LHM19,HLRM20,MRHLM21,HMWH22}, i.e. $L = 19$ a.u., $a_{+1} = 3.1$ a.u., $a_{-1} = 4.0$ a.u., $a_f = 5.0$ a.u., $\omega = 0.1$ a.u., $\lambda = 0.005$ a.u., and test the performance of standard (Eh, SH and SHEDC) and EF-based independent (SHXF) and coupled (CTMQC) trajectory-based approaches against the exact reference. The top panel of Figure~\ref{fig:LHM19merge} shows the BO and polaritonic surfaces and the NACs for this system. We observe that the interaction with the quantized light modes inside the cavity induces crossing regions with large non-adiabatic couplings. For the exact dynamics an initial gaussian wavepacket centered at $R_0=-4$ a.u. and variance $\sigma=\frac{1}{2\sqrt{2.85}}$ is launched on the second polaritonic surface. The time-dependent Schr\"odinger equation is solved in a three dimensional grid using the split-operator method~\cite{FFS82,BS93}. For the trajectory-based simulations $N_{tr}=$2000 Wigner-distributed trajectories, sampled from the same distribution, are run starting on the second polaritonic surface. The time-step used in the exact and trajectory-based simulations is $dt=0.024$ fs ($0.1$ a.u.).


Panels b) and c) of Figure~\ref{fig:LHM19merge} show the population dynamics. Let us first consider the exact dynamics. The initial wavepacket is launched in a region of strong coupling with the third state.  In the first 10 fs around 30\% of the population gets transferred to the 3rd polaritonic state. The wavepacket component created on the 3rd surface splits with part of the wavepacket moving to the left getting trapped in the well at $R \sim -4$ a.u., and part of the wavepacket moving to the right encountering at 15 fs the avoided crossing located at $R \sim -2$ a.u. ($d_{34}$) transferring population to state 4.  On the other hand, the wavepacket evolving on the second surface after the first interaction region reaches at 20 fs the second avoided crossing located at $R \sim -2$ a.u. ($d_{12}$) yielding partial population transfer to the lowest polaritonic state. Both of these states have a large electronic character, and the dynamics of this part of the wavepacket is similar to the cavity-free case~\cite{LHM19,MRHLM21}. 

The middle panel of Figure~\ref{fig:LHM19merge} shows the electronic populations and fraction of trajectories obtained with Eh, SH and SH-EDC. Taking a look at the electronic populations, Eh and SH overestimate both the first $2\rightarrow 3$ population transfer and the second $3\rightarrow 4$ population transfer but overall the trends are correct. The fraction of trajectories of SH is a bit closer to the exact results yielding poor internal consistency. On the other hand, with the decoherence correction, SHEDC gives improved population behavior and partially cures the internal inconsistency of SH.

The lowest panel of Figure~\ref{fig:LHM19merge} shows two of the EF-based MQC methods that are, in some sense, at different extremes:  CTMQC and SHXF. CTMQC slightly overestimates all population transfers, whereas the electronic populations in SHXF slightly underestimates them. The fraction of trajectories in SHXF follows the exact trend very closely. Overall, both SHEDC and SHXF capture decoherence well as we expected from a situation where nonadiabatic effects involve predominantly pair-wise interactions between the polaritonic states, as expected from the top panel of Figure~\ref{fig:LHM19merge}. 

\begin{figure}
	\includegraphics[width=0.45\textwidth]{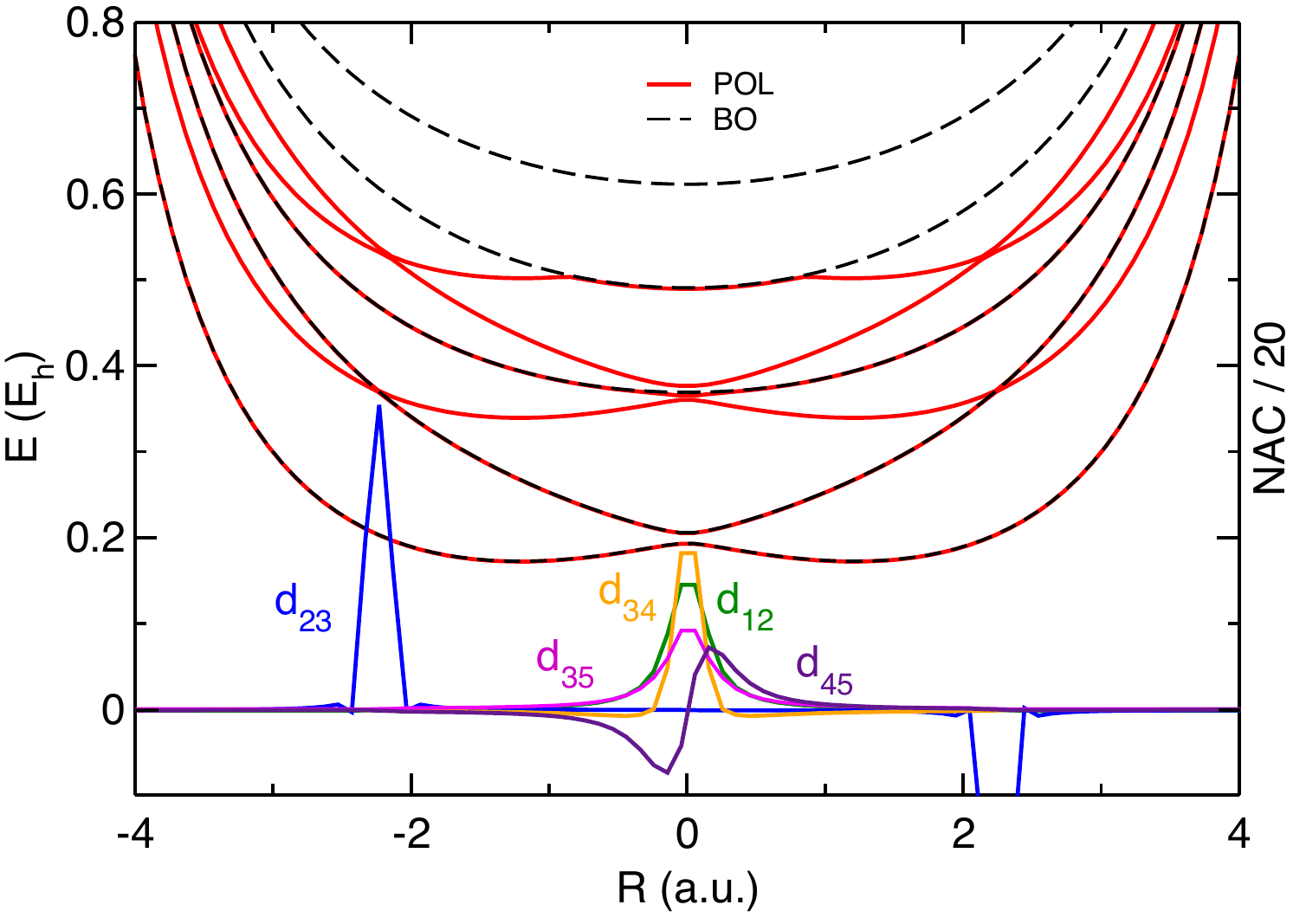}
	\caption{BO (cavity-free), polaritonic surfaces and NACs ($d_{ij}$) for a symmetric Shin-Metiu model, Eq.~(\ref{eq:SM}), with $L = 10$ a.u., $a_{\pm1} =$ 1.5 a.u., $a_f = 2.5$ a.u., in a cavity with frequency $\omega= 0.17$ a.u and light-matter coupling strength $\lambda = 0.01$ a.u.}
 \label{fig:sympes}
\end{figure}

Having tested the extension of the EF-based MQC methods to the polaritonic system studied in Ref.~\cite{LHM19,HLRM20,MRHLM21,HMWH22}, we now turn to investigate multi-state interactions. We choose a symmetric Shin-Metiu model, Eq.~(\ref{eq:SM}) , with parameters  $L = 10$ a.u., $a_{\pm1} =$ 1.5 a.u., $a_f = 2.5$ a.u. a cavity frequency $\omega= 0.17$ a.u and light-matter coupling strength
$\lambda = 0.01$ a.u. The polaritonic and BO (cavity-free) surfaces and the NACs are plotted in Fig. \ref{fig:sympes}. The chosen set of parameters results in a very different situation to our previous example, where now a a three-way avoided crossing involving  polaritonic states 3,4
and 5 located is induced around $R=0$ a.u. There are also avoided crossing between states 1 and 2 around $R=0$ a.u. and between states 2 and 3 at $R\sim\pm2$ a.u. We test the performance of independent and coupled EF-based trajectory methods against traditional schemes Eh and SH(-EDC). For the dynamics an initial Gaussian nuclear wavepacket with variance $\sigma=\frac{1}{2\sqrt{2.85}}$ centered around $R=-1.0$ a.u. is launched on the 4th polaritonic state. 


\begin{figure}
\includegraphics[width=0.45\textwidth]{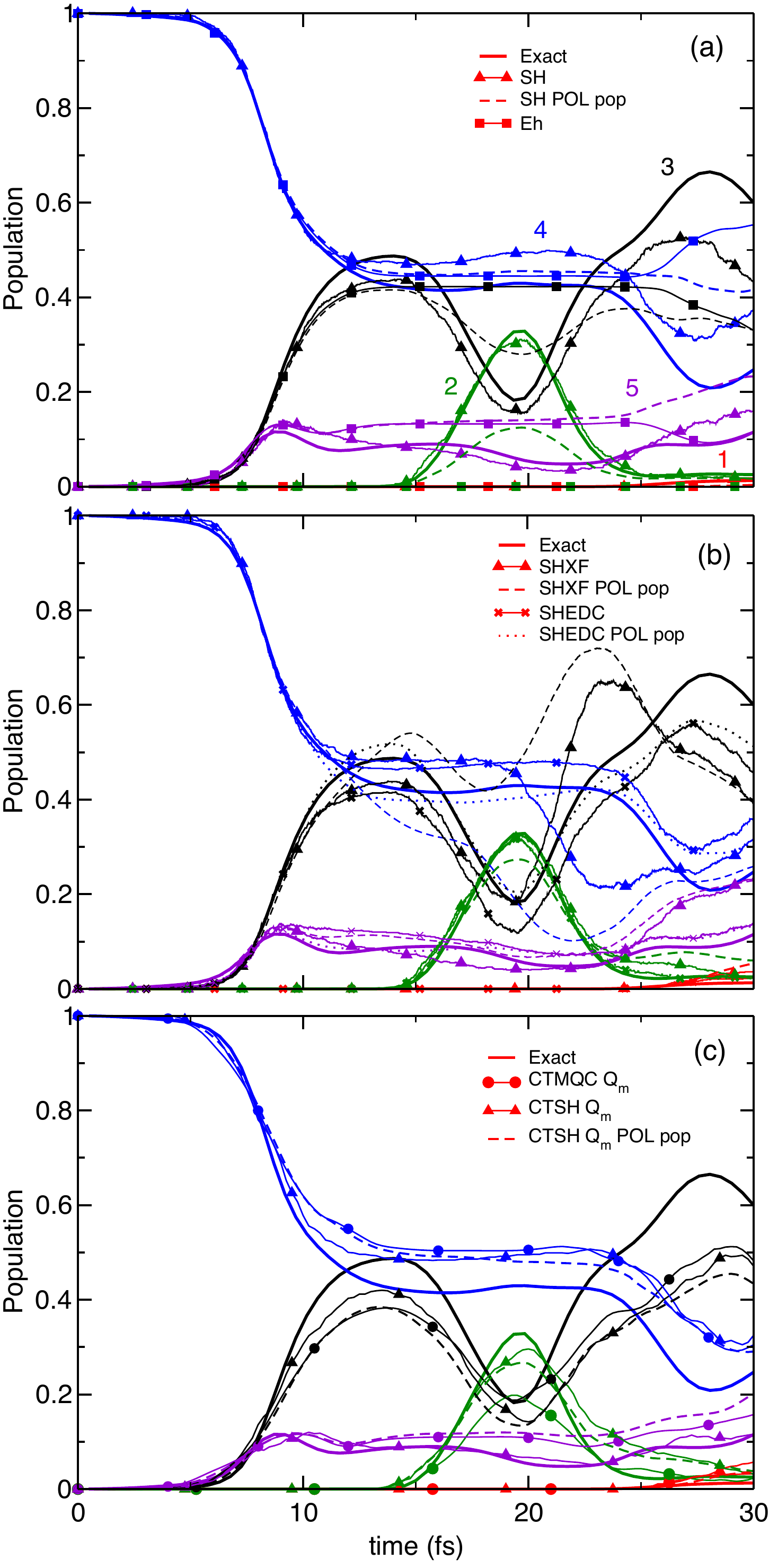}
	\caption{Population dynamics in the symmetric Shin-Metiu polaritonic model compared with exact calculations (solid lines). Red, green, black, blue and purple lines correspond to states 1 to 5, respectively. For the SH-based calculations, the populations are shown as fraction of trajectories (solid lines) and electronic populations (dashed lines). (a) Eh and SH. (b) SHXF and SHEDC. (c) CTMQC populations and CTSH with $Q_m$.} 
 \label{fig:sym-dyn}
\end{figure}

Figure \ref{fig:sym-dyn} shows the population dynamics with SH, Eh, SHEDC, SHXF, CTSH and CTMQC together with the exact reference. First, we take a look at the exact dynamics. In the first 10 fs, the nuclear wavepacket reaches the three-state avoided crossing region and we observe simultaneous population transfer to polaritonic states 5 and 3 at around 5 fs. The wavepacket component on the third polaritonic surfaces passes through the coupling region, located at $R\simeq 2.0$ a.u., at around 15 fs and transfers population to the second polaritonic state. The wavepacket component that remained on the fourth polaritonic surface at the three-state avoided-crossing gets reflected and reaches again the three-way avoided crossing region at 25 fs, transferring population to states 5 and 3. This situation, which involves dynamics through a multi-state intersection, resembles the case of the uracil radical cation that we saw in the previous section and we see that the population trends in the first 10 fs show similar transfer behavior in the two systems. Panel a) shows the dynamics obtained with standard SH and Eh. We observe that, the fraction of trajectories of SH reproduces the exact dynamics reasonably well, slightly underestimating the population transfer in the 3-state interaction region. The internal consistency is quite poor with the electronic populations greatly underestimating the population of state 2 and missing almost completely the population transfer from state 4 to states 3 and 5 ocurring at 25 fs. Eh on the other hand, misses completely the population transfer from state 3 to state 2 at 20 fs and predicts population transfer from state 3 and 5 to state 4 at around 25 fs.

We now take a look at the decoherence corrected SH methods SHEDC and SHXF, plotted in panel b) of Fig.~\ref{fig:sym-dyn}. We observe that SHEDC reproduced the exact dynamics quite accurately, curing the large internal inconsistency of SH. SHXF, unlike in the uracil cation example, does not do a good job reproducing the dynamics. The fraction of trajectories roughly reproduces qualitatively the exact population trends, with an underestimate of the initial population transfer from the 4th state to the 3rd state and a too early second transfer between these states. The electronic populations, however, experience spurious transfer early on as evidenced from the population of the 4th state that keeps on decreasing after 12 fs where the exact populations plateau; only a very small part of the trajectory ensemble is in a region where couplings to the 4th state are appreciable. The internal consistency is quite poor and can be understood from the fact that, unlike the electronic populations, the fewest-switches hopping probability does not depend directly on the quantum momentum and is proportional to the NACs, and consequently the trajectories will not make hops away from interaction regions.~\cite{VAM22}. 
 
We turn now to coupled-trajectories EF-based methods on panel c) of Fig.~\ref{fig:sym-dyn}. We see that all these methods accurately capture the correct population behavior, slightly underestimating the population transfer in the 3-state crossing and underestimating the population transfer to state 2 at 20 fs. 

\begin{figure}
	\includegraphics[width=0.45\textwidth]{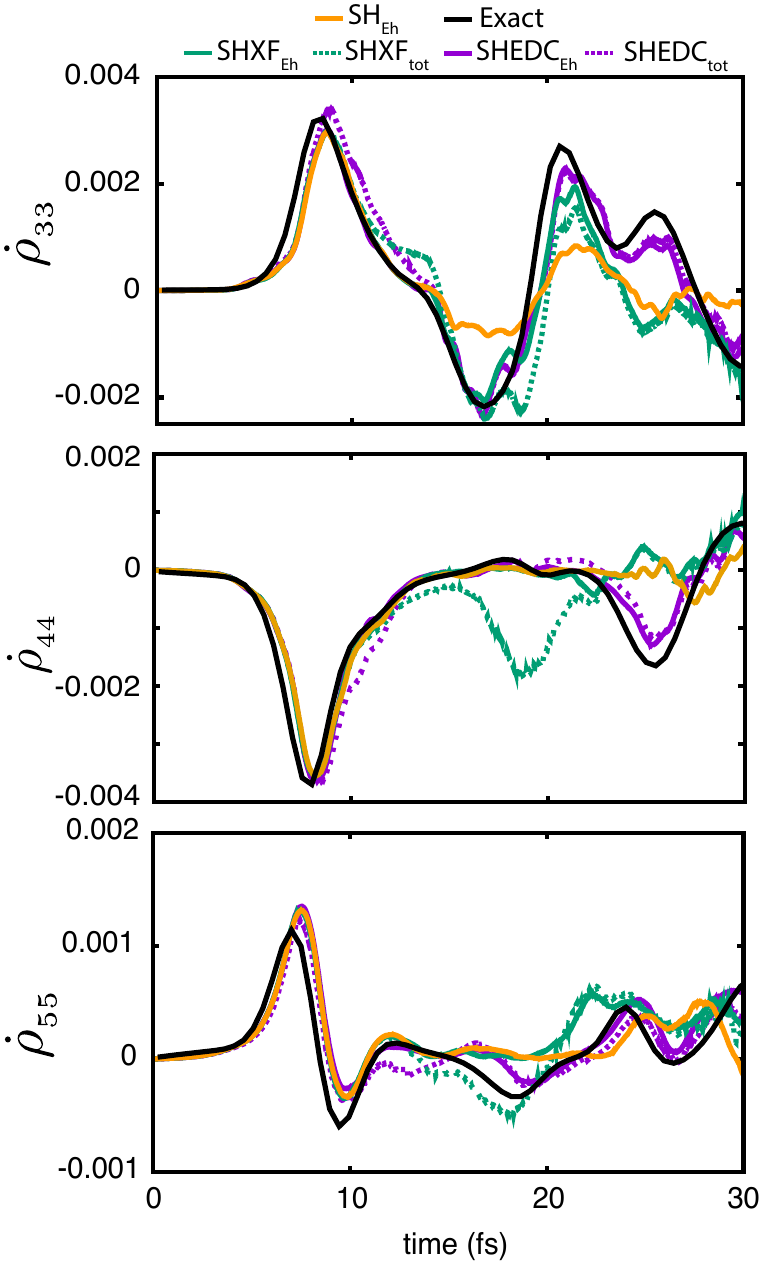}
	\caption{Total time derivative of the electronic populations ($\dot{\rho}_{kk}$) and Eh contribution ($\dot{\rho}_{Eh,kk}$) in SHEDC, SHXF (with all trajectories conserving the same total energy and trajectory fix) for states $k=3,4,5$ together with the exact reference. Subscripts $tot$ and $Eh$ indicate total time derivative and Eh-contribution respectively.}  \label{fig:sym-rhodot}
\end{figure}

Two main questions arise from these observations. The first is, 
why are SH and SHEDC reasonably accurate in capturing the dynamics through a 3-state intersection. The fail of traditional methods in describing a similar situation for the uracil cation, was attributed to the lack of quantum-momentum-driven electronic transitions. For a given electronic coefficient, the quantum-momentum term couples all electronic states, unlike the energy-decoherence correction term of SHEDC, that involves only a pair-wise interaction with the active state. This non-linear dependence on the coefficients in SHXF, is in contrast with the exponential rate of decay of SHEDC~\cite{VMM22}. A potential explanation to our first question could be that the quantum-momentum-driven transitions has a relatively minor effect on the dynamics, which are mostly dominated by the Eh terms, Eqs.~\ref{eq:el_eom}-\ref{eq:nuc_force}. This leads to our second question: Why does the quantum-momentum term in the electronic evolution of SHXF induce such a large error. To address whether our possible explanation to the first question could be correct, we compared, for the three states involved in the 3-state crossing, the trajectory-averaged value of the time-derivative of the electronic populations $\dot{\rho}_{kk}$ with the contribution coming from the Eh like term $\dot{\rho}_{Eh,kk}$ computed with all the SH-based schemes: SH, SHEDC and SHXF. Figure \ref{fig:sym-rhodot} shows these compared with the exact reference. We observe that for SHEDC, the Eh term dominates the full time-derivative in all three cases meaning that the decoherence correction term is relatively small throughout the dynamics when averaged over trajectories. The population rates are close to those of SH until about 15 fs, where they start to significantly differ after the first interaction region and where the decoherence correction begins to be active in SHEDC changing both the electronic and coupled nuclear dynamics compared to SH. Still, within SHEDC, even at later times the dominant term in the evolution of the populations is the Eh one. On the other hand, for SHXF we immediately observe that the contribution from the correction term to the EF-based term is more pronounced. Particularly for states 4 and 5, this term dominates the dynamics between 10-20 fs, which is consistent with the population transfer between these states lacking in the exact dynamics and in the other methods.

\begin{figure}	\includegraphics[width=0.45\textwidth]{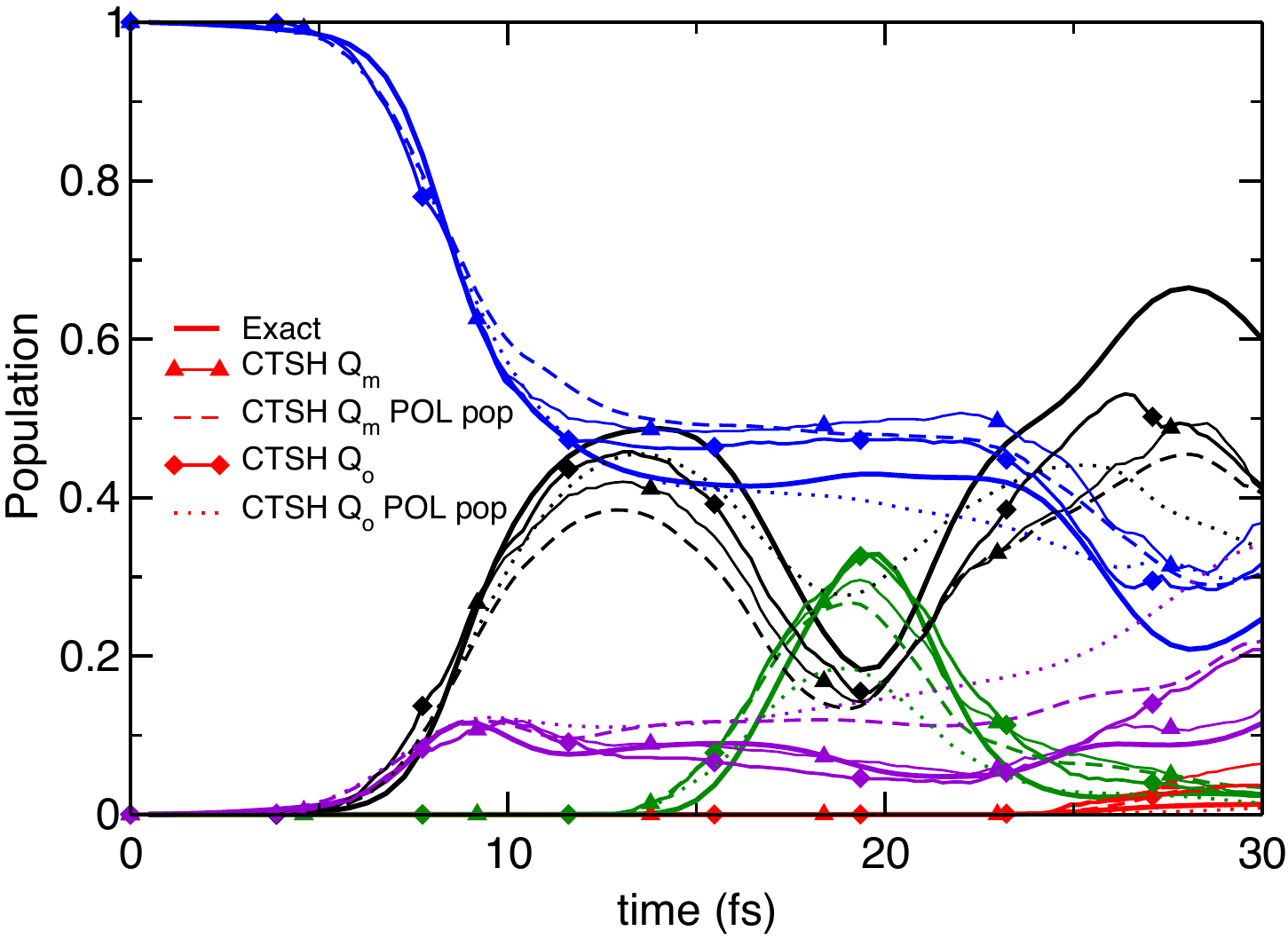}
	\caption{Population dynamics in the symmetric Shin-Metiu polaritonic model. (a) CTSH-$Q_m$ electronic populations (dashed lines) and SH fraction of trajectories (triangles). (b) CTSH-$Q_0$ electronic population (dotted lines) and SH fraction of trajectories (diamonds). \label{fig:comp-M}}
\end{figure}

The fact that the EF-based contribution in SHXF seems to act in regions where there should not be electronic population transfer, together with the good performance of the coupled-trajectories EF-based methods, suggests the answer for our second question lies in the way the quantum-momentum is computed via auxiliary trajectories. As discussed in Section \ref{sec:EF-based-mqc}, in SHXF, the quantum momentum is computed with $\textbf{Q}_0$ via auxiliary trajectories, which is liable to suffer from the spurious electronic population transfer~\cite{AMAG16,VMM22}. This leads to an active $\dot{\rho}_{EF,kk}$ term in non-interaction regions as we saw in Fig. \ref{fig:sym-rhodot}. The hopping probability however, does not depend on the quantum momentum directly and thus, will not suffer from spurious transfer, which leads to the large internal inconsistency observed in this case. On the other hand, CTMQC and CTSH use $\textbf{Q}_m$ by default, which imposes the condition of zero net contribution of the quantum-momentum term. To study this further, we ran CTSH approximating the quantum momentum via $\textbf{Q}_0$. Figure \ref{fig:comp-M} shows the electronic populations and fraction of trajectories for CTSH-$\textbf{Q}_0$ and CTSH-$\textbf{Q}_m$. Around 15 fs, CTSH with $Q_0$ shows spurious electronic population transfer between states 3 and 5, in contrast with the FT measure of the populations which follow the exact trend. Although both SHXF and CTSH-$Q_0$ suffer from spurious electronic transfer, the dynamics looks quite different with CTSH outperforming SHXF; this is in contrast to the uracil cation case where SHXF outperformed CTSH-$\textbf{Q}_0$. As discussed there, SHXF approximates the quantum momentum locally via $Q_o$ using one auxiliary trajectory per populated non-active state, while CTSH involves contributions from all the trajectories in the ensemble distributed among (all) active and non active states. 

\begin{figure}	\includegraphics[width=0.45\textwidth]{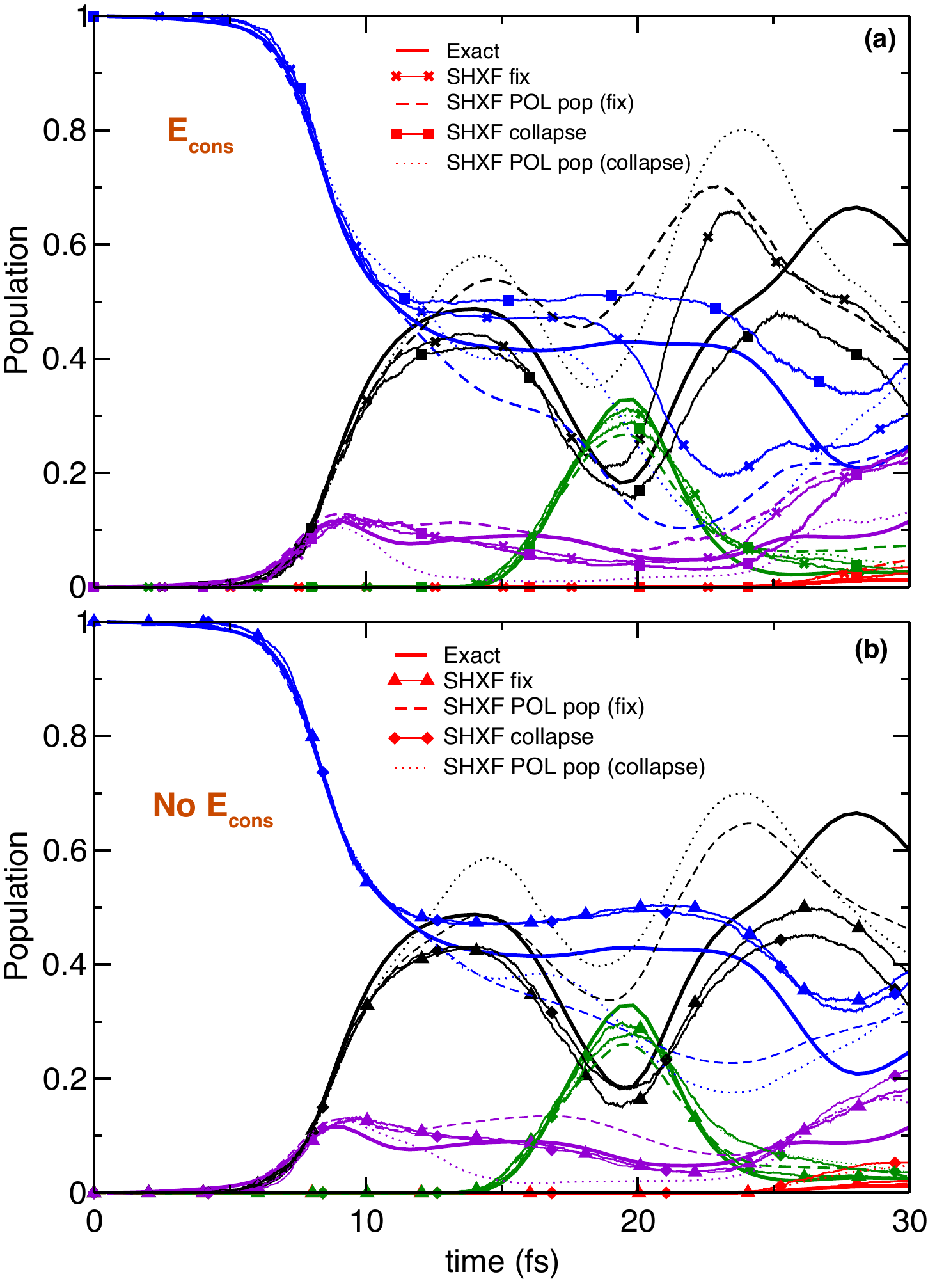}
\caption{SHXF population dynamics in the symmetric Shin-Metiu polaritonic model using different approaches to evolve the auxiliary trajectories: all auxiliary trajectories (a) are launched with the same total energy that they conserve, or (b) are launched with the same kinetic energy (and consequently different total energies). Each plot shows two options for auxiliary trajectories whose kinetic energy turns negative: velocity set to zero (denoted fix) until the coefficients are completely decohered, or coefficients of non-active states are collapsed to zero and the total electronic wavefunction is renormalized (denoted collapse).} \label{fig:econs}
\end{figure}

Finally, we explore the impact on the dynamics of two different aspects of the auxiliary propagation scheme: how to initialize the trajectories and how to deal with trajectories whose velocities turn complex due to requiring energy conservation. The auxiliary trajectories velocities are determined by isotropic rescaling of the real trajectory's velocity and they conserve total energy at each time-step during propagation as detailed in Appendix~\ref{app:A}. An essential aspect is to determine the initial auxiliary velocity, and thus its energy. The default in {PyUNIxMD} is to create the auxiliary trajectory with the same energy of the real trajectory, which implies that all auxiliary trajectories will have the same total energy. An alternative approach creates the auxiliary trajectory with the same kinetic energy of the real trajectory, which results in different total energies depending on which surface the auxiliary trajectory is launched on. A second key aspect is how to deal with trajectories for which the kinetic energy turns negative. The default implies fixing those trajectories until the coefficient on that state is fully decohered after which the auxiliary trajectories are killed anyway. Another approach is to destroy that auxiliary trajectory by collapsing the coefficient to zero. Fig.~\ref{fig:econs} shows the SHXF BO populations and fraction of trajectories obtained with different options in propagating the auxiliary trajectories. The upper panel of Fig.~\ref{fig:econs} (a) shows the situations where the energy of the  auxiliary trajectories is conserved throughout the propagation. We observe that applying the default trajectory fix induces a large spurious electronic transfer between states 4 and 3 at 10 fs and unphysical $4\rightarrow 3$ hops around 18 fs. When the auxiliary trajectory velocity is fixed, and the trajectory keeps propagating, the distance between the trajectory position and the quantum momentum center increases and, as we can see from Eq.(~\ref{eq:auxqmom}), so does the quantum momentum. On the other hand, collapsing the trajectory yields a more accurate fraction of trajectories, avoiding population transfer from state 4 to 3 at around 20 fs. Panel (b) shows SHXF dynamics when the  auxiliary trajectories are created with the same kinetic energy of the real trajectory and no energy conservation. We observe an improvement in both the fraction of trajectories and electronic populations over implementing energy conservation for the auxiliary trajectories. Whether we fix or collapse the auxiliary trajectory, the resulting fraction of trajectory trends follows the exact dynamics, although internal inconsistency still dominates. This inconsistency is more pronounced in states 3 and 4. However, for the uracil cation the standard SHXF implementation where all auxiliary trajectories conserve the same total energy, and where the velocity of auxiliary trajectories that cannot satisfy energy conservation is fixed yielded the most accurate dynamics.

\section{Conclusions and Outlook}
The EF-based MQC methods provide a first-principles description of electron-nuclear correlation effects such as decoherence and quantum-momentum driven electronic transitions, that are lacking in standard MQC schemes such as SH or Eh. The terms in the equations of motion involve the nuclear quantum momentum, related to the spatial variation of the nuclear density, and the accumulated forces related to the gradient of the phase of the electronic coefficients. They couple all electronic states in a non-linear way, which is a key difference to the traditional decoherence-corrected methods, that can significantly influence multi-state dynamics. In previous work, we found that the correction term derived from the EF approach yields dramatically improved agreement with reference quantum dynamics calculations in the uracil cation where more than two electronic state associated with a nuclear trajectory becomes occupied at a given time~\cite{VMM22}. Here, we extended our investigation and compared different flavors of EF-based MQC methods for such multi-state problems, involving either a three-state conical intersection or a three-way avoided crossing. Different EF-based MQC methods differ in whether they adopt an independent trajectory or coupled-trajectory algorithm, as well as in whether the EF term is adopted in both the electronic and nuclear equation or just the electronic equation, and in whether a surface-hopping or Ehrenfest "base" is used. The approximations going into the derivations of these methods can however lead to some of these methods violating of some physical constraints such as the condition that there should be zero net population transfer in regions of negligible coupling, and energy conservation. Our studies here have shown that when the dynamics occurs in a regime where these conditions are not important, all the EF-based methods provide an improvement over the traditional SH and Eh methods, in some cases quite significantly. When they are important, the EF-based methods that respect these conditions perform well. 

In Sec.~\ref{sec:uracil} we found that the EF-based methods all provide a qualitative improvement over the traditional methods. When used in an independent-trajectory framework, SHXF performed the best, outperforming the calculation where the EF term is kept in both the electronic and nuclear equation; this was conjectured to be perhaps a consequence of energy non-conservation in the latter approach while SHXF satisfies energy conservation. SHXF in principle violates the condition of zero net transfer in regions of zero NAC, but the dynamics shown apparently did not reach this regime. This was also verified by the closeness of the coupled-trajectory method with the original and modified definitions of quantum momentum  (CTSH-$\mathbf{Q}_0$ and CTSH-$\mathbf{Q}_m$); the former is not guaranteed to satisfy this condition while the latter is. Because the coupled-trajectory scheme uses non-local information from all the trajectories in the ensemble to approximate the quantum momentum, while the auxiliary trajectories of independent-trajectory approaches are more local, the CTSH results differed from the more accurate SHXF.. When used with coupled-trajectories, the  best performance was achieved when the EF terms are kept in both the electronic and nuclear evolution (CTMQC). The agreement with the reference MCTDH was strikingly good in the early evolution when the initial electronic was chosen to be a pure state for each nuclear trajectory reflecting the initial populations of the MCTDH state, rather than running a mixed state with each nuclear trajectory associated with a single electronic state, distributed according to the MCTDH initial populations. The pure state meant that the EF terms were effective in inducing population transfer from the very start, accurately capturing the initial decay, in contrast to when a mixed state was used. Energy non-conservation may be a factor for the less good agreement at later times.

While the violation of the two exact conditions did not adversely affect the uracil cation dynamics shown, they did affect our second example of multi-state dynamics (Sec 4). In particular, while the higher dimensionality of the model and the topology of the energy landscape in the uracil cation results in trajectories being in the vicinity of NAC regions throughout the duration of the dynamics studied, this was not the case for the one-dimensional polaritonic model displaying the three-state avoided crossing. In this case, the use of coupled-trajectory methods with the modified definition of quantum momentum that respects this condition, was key to capturing accurate dynamics. Further, imposing zero net population transfer is essential to correct the internal inconsistency inherent to SH schemes in systems where spurious population transfer dominates, as we observed in the symmetric polaritonic system. Interestingly, despite the three-state populations behaving initially similar to the uracil-cation case, where we argued that EF-based methods should provide important corrections over the traditional methods, we found for this case that the traditional SHEDC method worked well because the dynamics was dominated by the traditional terms in all the approaches. 

 Overall, EF-based MQC approaches tend to exhibit improved performance compared to traditional MQC methods when used within the coupled-trajectory framework with the modified definition of the quantum momentum, or when used with independent-trajectories framework with SH in cases of dynamics which remain within NAC regions. They arise as a promising and powerful tool for studying complex dynamics in molecules involving multiple electronic states and nonadiabatic events. Coupled trajectory methods can yield more accurate results emerging from the inclusion of coupling terms, albeit at a higher computational cost. Conversely, independent trajectory EF-based approaches provide improved predictions compared to traditional MQC methods at a similar computational expense, making them advantageous for investigating high-dimensional systems. We found a particularly interesting influence on the initial state choice: when the initial electronic character involves more than one BO state, an incoherent statistical mixture is usually chosen in classical trajectory methods such as SH, however in CT approaches, a more faithful representation of the initial state is a pure state where each trajectory in the ensemble is associated with non-zero electronic coefficients on each populated state. This has a dramatic effect on the dynamics, since the EF term is turned on from the very start. We found that only when this is done, is the initial dynamics very accurately captured. Current investigations are ongoing to explore the impact of different alternatives to build the initial state including the effect of phases on the electronic coefficients. 
Also underway is a deeper analysis of the energy non-conservation of CTMQC; this was relatively small in the cases studied here, but is not guaranteed to be small in general~\cite{VM23,HM22}. 
Ongoing developments of EF-based schemes that robustly satisfy exact conditions offer a promising reliable approach for coupled electron-nuclear dynamics. 

\appendix
\section{Auxiliary propagation schemes}
\label{app:A}
In the standard approach within the independent trajectories EF-based methods, the auxiliary trajectories used to locally approximate the quantum momentum, are launched as $\dulR_{\,k}^{(\alpha)}(t')=\dulR^{(\alpha)}(t')$ on non-active BO surfaces ($k\neq l$) that become populated at time $t'$, and follow uniform velocity motion during each time interval $[t',t'+\Delta t]$. Their velocities during propagation are determined by isotropic velocity rescaling of $\dot{\dulR}_{\,\nu }^{(\alpha)}(t')$, namely $\dot{\dulR}_{\,k,\nu }^{(\alpha)}(t')=\eta\dot{\dulR}_{\nu }^{(\alpha)}(t')\,\,;\,\, \eta\in\mathbb{R}$ enforcing energy conservation at each time-step:
\begin{equation}
\sum_\nu\frac{1}{2}M_\nu\dot{\mathbf{R}}_{\,k,\nu }^{(\alpha)\,2}(t+\Delta t)+E^{(\alpha)}_{k}(t+ \Delta t)=\sum_\nu\frac{1}{2}M_\nu\dot{\mathbf{R}}_{\,k,\nu }^{(\alpha)\,2}(t)+E^{(\alpha)}_{k}(t)\,. 
\label{eq:auxecons}
\end{equation}
Hence the auxiliary trajectory velocity reads:
\begin{equation}
\dot{\dulR}_{\,k,\nu }^{(\alpha)}(t+\Delta t)=\sqrt{\frac{E^{(\alpha)}_{k}(t)+\sum_\nu\frac{1}{2}M_\nu\dot{\mathbf{R}}_{\,k,\nu }^{(\alpha)\,2}(t)-E^{(\alpha)}_{k}(t+\Delta t)}{\sum_\nu\frac{1}{2}M_\nu\dot{\mathbf{R}}_\nu^{(\alpha)\,2}(t+\Delta t)}}\,\dot{\dulR}_{\nu }^{(\alpha)}(t)\,.
\label{eq:uvr2} 
\end{equation}
Their initial velocities can be set in two ways. The standard approach launches these with the same total energy of the real trajectory
\begin{equation}
E^{(\alpha)}=\sum_\nu\frac{1}{2}M_\nu\dot{\mathbf{R}}_\nu^{(\alpha)\,2}(t)+E^{(\alpha)}_{l}(t)=\sum_\nu\frac{1}{2}M_\nu\dot{\mathbf{R}}_{\,k,\nu }^{(\alpha)\,2}(t)+E^{(\alpha)}_{k}(t)\,. 
\label{eq:auxecons}
\end{equation}
which means the auxiliary trajectory velocity reads:
\begin{equation}
\dot{\dulR}_{\,k,\nu }^{(\alpha)}(t)=\sqrt{1-\frac{\Delta E^{(\alpha)}_{l,k}(t)}{\sum_\nu\frac{1}{2}M_\nu\dot{\mathbf{R}}_\nu^{(\alpha)\,2}(t)}}\,\dot{\dulR}_{\nu }^{(\alpha)}(t)\,.
\label{eq:uvr} 
\end{equation}
In this way, the algorithm sets identical total energy for all auxiliary trajectories $E_{aux,k}^{(\alpha)}(t)=E^{(\alpha)}$. An alternative approach consists on launching the auxiliary trajectories with the same kinetic energy of the real trajectory, i.e.,  $\dot{\dulR}_{\,k,\nu }^{(\alpha)}(t')=\dot{\dulR}_{\nu }^{(\alpha)}(t')$. Therefore all auxiliary trajectories have different total energies $E_{aux,k}^{(\alpha)}(t)\neq E_{aux,j}^{(\alpha)}(t)\neq E^{(\alpha)}$. 
From Eq.~(\ref{eq:uvr2}) we can see that the velocity of the auxiliary trajectory could become complex, situation that represents a a classically forbidden region as the auxiliary trajectory penetrates a potential energy barrier with lower total energy $E_{aux}^{(\alpha)}(t)<E_l^{(\alpha)}(t+\Delta t)$.  The original SHXF algorithm deals with such situations by setting the auxiliary trajectory velocity to zero, $\dot{\dulR}_{\,l}^{(\alpha)}(t')=0$, until the population of its corresponding non-active state is (numerically) zero and the auxiliary trajectory is destroyed~\cite{HLM18}. An alternative approach, 
borrowed from the branching corrected surface hopping (BCSH) method~\cite{XW19}, consists on collapsing the coefficient of the non-active state to zero, and renormalizing the electronic wavefunction. The underlying idea is to avoid situations where a reflecting wavepacket leads to multiple wavepacket components on a given surface. In these cases, the BH expansion of the instantaneous time-dependent electronic wavefunction along a trajectory 
$\Phi^{(\alpha)}(\dulr,t)=\sum_lC_l^{(\alpha)}(t)\phi_l^{(\alpha)}(\dulr)$
breaks down.

\section*{Conflicts of interest}
There are no conflicts to declare.

\section*{Acknowledgements}
Financial support from the National Science Foundation Award CHE-2154829 and from the Department
of Energy, Office of Basic Energy Sciences, Division of Chemical
Sciences, Geosciences and Biosciences under Award No. DE-SC0020044 and the Computational Chemistry
Center: Chemistry in Solution and at Interfaces funded by the
U.S. Department of Energy, Office of Science Basic Energy
Sciences, under Award DE-SC0019394 as
part of the Computational Chemical Sciences Program is gratefully acknowledged. Supplement funding for this project was provided by the Rutgers University at Newark Chancellor's Research Office. 



\bibliography{ref_na.bib} 
\bibliographystyle{rsc} 

\end{document}